%% file: preprint.tex
\documentclass[prd,floatfix,showpacs,preprintnumbers]{revtex4}
\usepackage{graphicx}
\usepackage{epsf}
\usepackage{epsfig}
\usepackage{color}
\usepackage{bm}
\usepackage{longtable}

\newcommand{\mco}{\multicolumn}

\newcommand{\be}{\begin{equation}}
\newcommand{\ee}{\end{equation}}
\newcommand{\bea}{\begin{eqnarray}}
\newcommand{\eea}{\end{eqnarray}}

\newcommand{\Ampl}{{\mathcal A}}
\let \bar=\overline
\let \to=\rightarrow

\newcommand{\EIPWA}{MIPWA}
\newcommand{\FD}[1]{\ensuremath{{\cal F}_{\sst D}^{\sst #1}}}
\newcommand{\rdpl}{\ensuremath{r_{\sst D}}}
\renewcommand{\FR}[1]{\ensuremath{{\cal F}_{\sst R}^{\sst #1}}}
\newcommand{\rres}{\ensuremath{r_{\sst R}}}
\newcommand{\piA}{\ensuremath{\pip_{\sst A}}}
\newcommand{\piB}{\ensuremath{\pip_{\sst B}}}
\newcommand{\SA}{\ensuremath{s}}
\newcommand{\SAA}{\ensuremath{s_{\sst A}}}
\newcommand{\SAB}{\ensuremath{s_{\sst B}}}
\newcommand{\SAk}{\ensuremath{s_k}}
\newcommand{\bwph}{\ensuremath{\phi_{\sst BW}}}
\newcommand{\coeff}[1]{\ensuremath{B_{\sst #1}}}
\newcommand{\brat}[1]{\ensuremath{F_{\sst #1}}}
\newcommand{\mR}{\ensuremath{m_{\sst R}}}
\newcommand{\pR}{\ensuremath{p_{\sst R}}}
\newcommand{\GamR}{\ensuremath{\Gamma_{\sst R}}}
\newcommand{\swave}{\ensuremath{ S\hbox{-wave}}}
\newcommand{\pwave}{\ensuremath{ P\hbox{-wave}}}
\newcommand{\dwave}{\ensuremath{ D\hbox{-wave}}}
\newcommand{\sdash}{\ensuremath{ S\hbox{-}}}
\newcommand{\pdash}{\ensuremath{ P\hbox{-}}}
\newcommand{\ddash}{\ensuremath{ D\hbox{-}}}
\newcommand{\KAs}{\ensuremath{\kappa}}
\newcommand{\KBs}{\ensuremath{K_0^*(1430)}}
\newcommand{\KAp}{\ensuremath{K^*(892)}}
\newcommand{\KBp}{\ensuremath{K_1^*(1410)}}
\newcommand{\KCp}{\ensuremath{K_1^*(1680)}}
\newcommand{\KAd}{\ensuremath{K_2^*(1430)}}

\newcommand{\NDF}{\hbox{\small NDF}}
\newcommand{\chidof}{\ensuremath{\chi^2/\NDF}}

\newcommand{\gammaz}{\ensuremath{\gamma_{0}}}
\newcommand{\pvec}{\ensuremath{{\cal P}}}
\newcommand{\asymm}{\ensuremath{\alpha}}
\newcommand{\BW}{\ensuremath{{\cal W}}}
\newcommand{\three}{\ensuremath{Q}}

\input mytex.tex

\input babar.tex

\begin{document}

\title{Model Independent Measurement of \swave\ $K^{-}\pi^{+}$ 
 systems using $\Dp\to K\pi\pi$ Decays 
 from Fermilab E791.}
 \author{
    E.~M.~Aitala,$^9$
       S.~Amato,$^1$
    J.~C.~Anjos,$^1$
    J.~A.~Appel,$^5$
       D.~Ashery,$^{14}$
       S.~Banerjee,$^5$
       I.~Bediaga,$^1$
       G.~Blaylock,$^8$
    S.~B.~Bracker,$^{15}$
    P.~R.~Burchat,$^{13}$
    R.~A.~Burnstein,$^6$
       T.~Carter,$^5$
    H.~S.~Carvalho,$^{1}$
    N.~K.~Copty,$^{12}$
    L.~M.~Cremaldi,$^9$
       C.~Darling,$^{18}$
       K.~Denisenko,$^5$
       S.~Devmal,$^3$
       A.~Fernandez,$^{11}$
    G.~F.~Fox,$^{12}$
       P.~Gagnon,$^2$
       C.~Gobel,$^1$
       K.~Gounder,$^9$
    A.~M.~Halling,$^5$
       G.~Herrera,$^4$
       G.~Hurvits,$^{14}$
       C.~James,$^5$
    P.~A.~Kasper,$^6$
       S.~Kwan,$^5$
    D.~C.~Langs,$^{12}$
       J.~Leslie,$^2$
       B.~Lundberg,$^5$
       J.~Magnin,$^1$       
       A.~Massafferri,$^1$
       S.~MayTal-Beck,$^{14}$
       B.~Meadows,$^3$
       J.~R.~T.~de~Mello~Neto,$^1$
       D.~Mihalcea,$^7$
    R.~H.~Milburn,$^{16}$
    J.~M.~de~Miranda,$^1$
       A.~Napier,$^{16}$
       A.~Nguyen,$^7$
    A.~B.~d'Oliveira,$^{3,11}$
       K.~O'Shaughnessy,$^2$
    K.~C.~Peng,$^6$
    L.~P.~Perera,$^3$
    M.~V.~Purohit,$^{12}$
       B.~Quinn,$^9$
       S.~Radeztsky,$^{17}$
       A.~Rafatian,$^9$
    N.~W.~Reay,$^7$
    J.~J.~Reidy,$^9$
    A.~C.~dos Reis,$^1$
    H.~A.~Rubin,$^6$
    D.~A.~Sanders,$^9$
 A.~K.~S.~Santha,$^3$
 A.~F.~S.~Santoro,$^1$
       A.~J.~Schwartz,$^{3}$
       M.~Sheaff,$^{17}$
    R.~A.~Sidwell,$^7$
    A.~J.~Slaughter,$^{18}$
     M.~D.~Sokoloff,$^3$
       J.~Solano,$^1$
    N.~R.~Stanton,$^7$
    R.~J.~Stefanski,$^5$  
       K.~Stenson,$^{17}$ 
    D.~J.~Summers,$^9$
       S.~Takach,$^{18}$
       K.~Thorne,$^5$
    A.~K.~Tripathi,$^{7}$
       S.~Watanabe,$^{17}$
 R.~Weiss-Babai,$^{14}$
       J.~Wiener,$^{10}$
       N.~Witchey,$^7$
       E.~Wolin,$^{18}$
    S.~M.~Yang,$^7$
       D.~Yi,$^9$
       S.~Yoshida,$^7$
       R.~Zaliznyak,$^{13}$ and
       C.~Zhang$^7$ \\
\begin{center}
Fermilab E791 Collaboration \\
$^1$ Centro Brasileiro de Pesquisas F{\'\i}sicas, Rio de Janeiro, Brazil\\
$^2$ University of California, Santa Cruz, California 95064\\
$^3$ University of Cincinnati, Cincinnati, Ohio 45221\\
$^4$ CINVESTAV, Mexico City, Mexico\\
$^5$ Fermilab, Batavia, Illinois 60510\\
$^6$ Illinois Institute of Technology, Chicago, Illinois 60616\\
$^7$ Kansas State University, Manhattan, Kansas 66506\\
$^8$ University of Massachusetts, Amherst, Massachusetts 01003\\
$^9$ University of Mississippi-Oxford, University, Mississippi 38677\\
$^{10}$ Princeton University, Princeton, New Jersey 08544\\
$^{11}$ Universidad Autonoma de Puebla, Puebla, Mexico\\
$^{12}$ University of South Carolina, Columbia, South Carolina 29208\\
$^{13}$ Stanford University, Stanford, California 94305\\
$^{14}$ Tel Aviv University, Tel Aviv, Israel\\
$^{15}$ Box 1290, Enderby, British Columbia, V0E 1V0, Canada\\
$^{16}$ Tufts University, Medford, Massachusetts 02155\\
$^{17}$ University of Wisconsin, Madison, Wisconsin 53706\\
$^{18}$ Yale University, New Haven, Connecticut 06511\\[12pt]
        W.~M.~Dunwoodie \\
        SLAC., Stanford, California 94305
\end{center}
}

 \noaffiliation
 \vspace{10pt}
 \date{\today}
 \begin{abstract}
 \input{abstract.tex}
 \end{abstract}
 \preprint{FERMILAB-PUB-05-336-E}
 \preprint{UCHEP-05-03}
 \maketitle

\section{\bf Introduction}
\label{sec:intro}
\input{intro.tex}

\section{\bf The E791 Data}
\label{sec:dalitzplot}
\input{dalitzplot}

\section{\bf Formalism}
\label{sec:method}
\input{method.tex}

\section{\bf \EIPWA\ of the $\Km\pip$ \swave}
\label{sec:swave}
\input{swave_fit.tex}
\section{Comparison with an isobar model fit}
\label{sec:kappatest}
\input{kappatest.tex}
\section{\bf Comparison of \EIPWA\ with Elastic Scattering}
\label{sec:elastic}
\input{elastic.tex}

\section{\bf Systematic uncertainties}
\label{sec:systematic}
\input{systematic.tex}

\section{\bf Summary and conclusions}
\label{sec:summary}
\input{summary.tex}

\begin{acknowledgments}
We wish to thank members of the LASS collaboration for making 
their data available to us.
We gratefully acknowledge the assistance of the staffs of Fermilab
and of all the participating institutions.  This research was
supported by the Brazilian Conselho Nacional de 
Desenvolvimento Cient\'{\i}fico e Tecnol\'{o}gico,
CONACyT (Mexico), FAPEMIG
(Brazil), the Israeli Academy of Sciences and Humanities,
the U.S. Department of Energy, the U.S.-Israel
Binational Science Foundation, and the U.S. National Science 
Foundation.  Fermilab is operated by the Universities Research
Association for the U.S. Department of Energy.
\end{acknowledgments}

\appendix



\section{Limitations and Technicalities of the method}
\label{sec:ambiguities}
\input{ambiguities.tex}

\bibliography{preprint}

\end{document}

%% file: mytex.tex
\newcommand{\sst}{\scriptscriptstyle}

\newcommand{\etc}{{\sl etc~}}

\newcommand{\half}{\ensuremath{{1\over 2}}}
\newcommand{\z}{\ensuremath{_{\circ}}}

\newcommand{\MeVcc}{\ensuremath{\hbox{MeV}/c^2}}
\newcommand{\GeVcc}{\ensuremath{\hbox{GeV}/c^2}}

%% file: babar.tex
\let\mathrm=\rm

\newcommand{\KS}{K^{0}_s}

\newcommand{\Dz}{D^{0}}

\newcommand{\pip}{\pi^+}
\newcommand{\pim}{\pi^-}
\newcommand{\Kp}{K^+}
\newcommand{\Km}{K^-}

\newcommand{\Dp}{D^+}

%% file: abstract.tex
 A model-independent partial-wave analysis
 of the \swave\ component of the $K\pi$
 system from decays of $D^{+}$ mesons to the three-body
 $\Km\pip\pip$ final state is described.  Data come from the
 Fermilab E791 experiment.  Amplitude measurements are made 
 independently
 for ranges of $\Km\pip$ invariant mass, and results
 are obtained below 825~\MeVcc, where previous measurements 
 exist only in two mass bins.
 This method of parametrizing a three-body decay amplitude
 represents a new approach to analysing
 such decays.  Though no model is required for the \swave,
 a parametrization of the relatively well-known reference 
 \pdash\ and \dwave s, optimized to describe the data used, is 
 required.  In this paper, a Breit-Wigner model is adopted to describe
 the resonances in these waves.
 The observed phase variation for the \sdash, \pdash\ and \dwave s
 do not match existing measurements of $I=\half$ $\Km\pip$ scattering 
 in the invariant mass range 
 in which scattering is predominantly elastic.
 If the data are mostly $I=\half$, this observation indicates
 that the Watson theorem, which requires these phases to have the
 same dependence on invariant mass, may not apply to these decays 
 without allowing for some interaction with the other pion.
 The production rate of $\Km\pip$ from these decays, if assumed to 
 be predominantly $I=\half$, is also found to have a significant
 dependence on invariant mass in the region above 1.25~\GeVcc.
%
 These measurements can provide a relatively model-free basis for 
 future attempts to determine which
 strange scalar amplitudes contribute to the decays.

%% file: intro.tex
Kinematics and angular momentum conservation in decays of ground state,
heavy-quark mesons to three
pseudoscalars strongly favor production of \swave\ systems.  These 
decays have therefore been regarded as a source of information on the
composition of the scalar meson (spin-parity $J^P=0^+$) spectrum.
Extracting this information has, however, been done in model-dependent
ways that can influence the outcome.  For the di-meson subsystems, 
vector and tensor resonances
are relatively well-understood, but, as larger samples of $D$ and $B$
meson decays become available, the correct modelling of the \swave\
contributions becomes an increasingly important factor in the task of
obtaining satisfactory fits to the data.  

Analyses typically use an isobar model formulation in which the decays 
are described by a coherent sum of a non-resonant three-body amplitude $NR$,
usually taken to be constant in magnitude and phase over the entire 
Dalitz plot, and a number of quasi two-body (resonance + bachelor) amplitudes
where the bachelor particle is one of the three final state products,
and the resonance decays to the remaining pair.
It is assumed that all resonant and $NR$ processes taking part in 
the decay are 
described by amplitudes that interfere, and have
relative phases and magnitudes determined by the decay of the parent 
meson.
In cases where all
three decay products are pseudoscalar ($P$) particles, angular
momentum conservation requires that the resonances produced are 
scalar (\swave), vector (\pwave), \etc.  For $D$ mesons, decays 
beyond \dwave\ are highly suppressed by the angular momentum barrier 
factor and can be neglected.

Within this formalism, the decays $\Dp\to\pim\pip\pip$ and
$\Dp\to\Km\pip\pip$
\cite{conjugate}
were once thought to require very large, constant $NR$ amplitudes
\cite{Frabetti:1994di, Anjos:1992kb, Frabetti:1997sx}.
Using larger samples, the Fermilab E791 collaboration found that a 
satisfactory description of these decays requires more structure.
By including \swave\ isobars,
$\sigma(500)\to\pip\pim$ in $\pim\pip\pip$
\cite{Aitala:2000xu}
and $\kappa(800)\to\Km\pip$ in $\Km\pip\pip$
\cite{Aitala:2002kr}, a much-improved modelling of the Dalitz plots
was achieved, and the need for a constant $NR$ term was much reduced
in each case.

The FOCUS collaboration, using an even larger
sample of $\Dp\to\pim\pip\pip$ decays, found an acceptable fit
\cite{Link:2003gb}
using a $K$\-matrix description of the \swave\ with no $\sigma(500)$
pole.  However, a parametrization of the $NR$ background was required
to achieve an acceptable fit.
The BaBar and Belle collaborations
\cite{Aubert:2005iz, Poluektov:2004mf, Abe:2005ct},
with the measurement of $CP$ violation parameters in
$B^-\to\Dz(\to\KS\pip\pim)\Km$ decays as their primary goal, introduce
$\sigma(500)$ and another $\sigma(1000)$ isobar in order to
obtain an acceptable description of the complex amplitude for the $D^0$
Dalitz plot.

The important issue of whether scalar particles $\sigma(500)$ and
$\kappa(800)$ exist is not convincingly settled.  Further
observations of these isobars were recently reported in $\pim\pip$
and $\Km\pip$ systems from $J/\psi$ decays
\cite{Ablikim:2004qn, Komada:2003gu}.
However, these results were modelled on variations of the simple
Breit-Wigner form for the states adopted in the cases cited.
Quite different descriptions are probably required, since such forms
are seen to contain poles below threshold
\cite{Gardner:2001gc}.
In a recent publication
\cite{Oller:2004xm}
the E791 data on $\Dp\to\pim\pip\pip$ and on the $\Dp\to\Km\pip\pip$
decays discussed here, are re-fitted using input from calculations
of $\pi\pi$ and $K\pi$ scattering that include constraints of
chiral perturbation theory, and that find both $\sigma$ and $\kappa$ 
poles
\cite{Oller:1997ti,Jamin:2000wn}.  
The fits obtained yield similar $\chi^2$ per degree of freedom to 
those in Refs.
~\cite{Aitala:2000xu} and \cite{Aitala:2002kr} 
where Breit-Wigners were used, but each resonance is considerably 
wider.

Ultimately, a less model-dependent analysis of the data should
help resolve the issue of the $\sigma$ and the $\kappa$.

Model-independent measurements of the energy dependence of these \swave\
amplitudes, particularly in the low invariant mass regions, where 
confusion is greatest, is therefore an important experimental goal.
Such a  Model-Independent Partial Wave Analysis (\EIPWA) is reported
here for the $\Km\pip$ system produced in $\Dp\to\Km\pip\pip$ decays.
One earlier measurement has been made for $\pim\pip$ systems from
$\Dp\to\pim\pip\pip$ decays
\cite{Bediaga:2004bc},
in which the ``amplitude difference'' (AD) method
\cite{Bediaga:2002au}
was employed.
This method can only be used when there exists a region of the Dalitz
plot that can be described by the sum of a single resonance and an
\swave\ amplitude that is to be measured.  
Interference of the resonance with this \swave\ introduces an asymmetry
in the distribution of the other invariant mass combinations
that can be measured at different values of invariant masses in the band.
As there is no such region in the Dalitz plot for the data reported here, 
this method is not used.

For the $\Km\pip$ system, the best results of an \EIPWA\ currently
available come from the LASS experiment
\cite{Aston:1987ir},
in which $\Km\pip$ scattering was studied for invariant masses
above 825~\MeVcc.  Below 825 \MeVcc, measurements have been made for
the mass bins 770-790~\MeVcc
\cite{Bingham:1972vy} and 700-760~\MeVcc
\cite{Estabrooks:1977xe}, though with less precision.
Information on the $K\pi$ \swave\ amplitude near or slightly above the
\KAp\ has been extracted by the BaBar collaboration in studies of $B$ 
decays to $J/\psi K\pi$
\cite{Aubert:2004cp},
and by FOCUS in semi-leptonic $D$ decays to $K\pi\ell\nu$
\cite{Link:2002ev},
the low mass region has not been covered in either case.

In this paper, we describe an \EIPWA\ in the mass range from $\Km\pip$
threshold up to 1.72~\GeVcc, the kinematic limit for
decays of $\Dp$ mesons to $\Km\pip\pip$ final states.
The amplitudes obtained for the \swave\ require no assumptions about
its dependence on invariant mass, though
they do rely on a model for the relatively well-understood \pdash\ and 
\dwave s.  As such, they should provide an
unbiased input for comparisons with theoretical models for scalar 
states.

This paper is organized as follows.
In the following section we present the data sample.  Next we describe
the method used to extract complex amplitudes from the \swave\ $\Km\pip$
system in a way that does not require a model for its dependence on
invariant mass.  In Section~\ref{sec:swave} this is applied to the sample 
of $\Dp\to\Km\pip\pip$ decays.  The amplitudes obtained are then compared, 
in Section~\ref{sec:kappatest}, with the \swave\ amplitude derived
from the Breit-Wigner isobar model fit that best represents the data.
This model, applied to these data, was presented in Ref.~\cite{Aitala:2002kr},
and includes a $\kappa(800)$ isobar.  In Section~\ref{sec:elastic} the
results of the \EIPWA\ are compared with amplitudes measured in $\Km\pip$ 
elastic scattering, and with the expectations of the Watson theorem
\cite{watson:1952ji},
whose applicability to weak hadronic decays has not previously been 
tested.  Systematic uncertainties are discussed in 
Sec.~\ref{sec:systematic}.  Finally, some conclusions are drawn.
In Appendix \ref{sec:ambiguities} we point out limitations, ambiguities 
and other technicalities inherent in this kind of analysis.

%% file: dalitzplot.tex
The analysis is based on a sample of $\Dp\to\Km\pip\pip$
candidates from Fermilab experiment E791.  The experiment is described
in detail in
Ref.~\cite{Aitala:1998kh}.
The same sample is used in this paper as the one described in
Ref.~\cite{Aitala:2002kr},
where an isobar model fit to these data was described.  The
selection process used in obtaining the sample is outlined
below, but more details are given in
Ref.~\cite{Aitala:2002kr}.

In this paper, \SA\ is used to denote the $\Km\pip$ squared 
invariant mass.  Where it is important to distinguish, the 
two pions (and their corresponding \SA\ values) are labelled,
respectively, $\piA$ and $\piB$ (\SAA\ and \SAB).
A clear peak in the $\Km\pip\pip$ invariant mass $M$
distribution is observed with 15,079 events in the mass range 
$1.810<M<1.890$ \GeVcc, of which 94.4\% are determined to be signal.
The major sources of background are incorrect three-body 
combinations (3.58\%), and reflections of $D_s^+\to\phi\pip$ and
$D_s^+\to\bar{K}^{*0}\Kp$ decays
(1.75\% and 2.61\%, respectively) in which a $K^+$ is incorrectly
identified as a $\pip$.  The probability density 
function (PDF) for these backgrounds over the Dalitz plot is obtained
from events in the sideband region of the $\Km\pip\pip$ invariant 
mass distribution and, for the second and third sources,
from a large sample of Monte Carlo (MC) simulated events.
An appropriately weighted combination of these three backgrounds is
determined from their distributions in $M$.  The
efficiency for reconstructing the $D^+$ decays (the signal) is also 
determined from the MC events.  It is described in this paper by a
function $\epsilon(\SAA, \SAB)$.

\subsection{E791 Dalitz Plot}

\begin{figure}[hbt]
 \centerline{%
 \epsfig{file=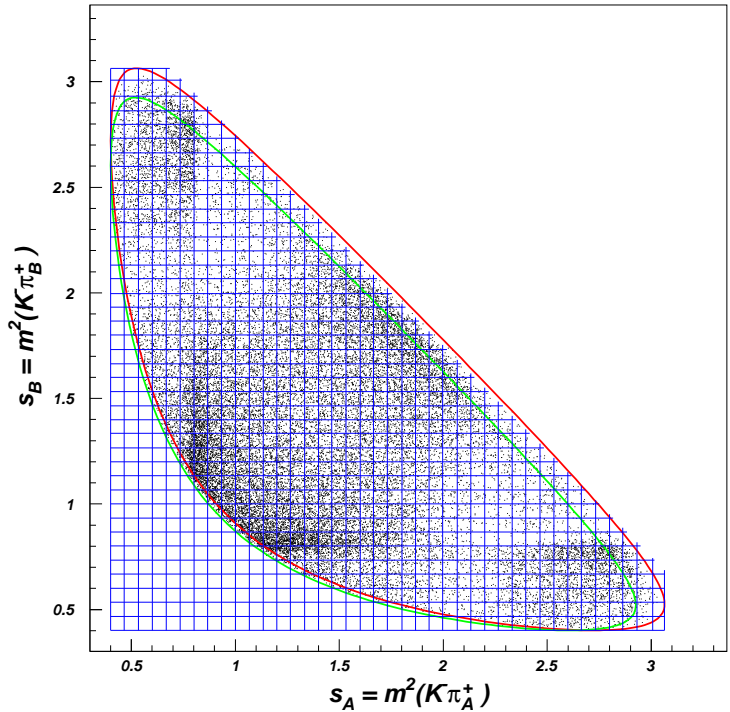,width=0.6\textwidth,angle=0}}
 \caption{Dalitz plot for $\Dp \to \Km\piA\piB$ decays.  The
  squared invariant mass \SAB\ of one $\Km\pip$ combination is
  plotted against $\SAA$, the squared invariant mass of the other
  combination.  The plot is symmetrized, each event
  appearing twice.  Lines in both directions indicate values equally
  spaced in squared effective mass at each of which the \swave\
  amplitude is determined by the \EIPWA\ described in section
  \ref{sec:method}.  Kinematic boundaries for the Dalitz plot are
  drawn for three-body mass values $M=1.810$ and $M=1.890$~\GeVcc,
  between which data are selected for the fits.
  \label{fig:dalitz_plot}}
\end{figure}

The symmetrized Dalitz plot for this sample is shown in Fig.
\ref{fig:dalitz_plot} where \SAA\ is plotted vs. \SAB\ (and the
converse).
A horizontal (and the symmetrized vertical) band corresponding
to the presence of the \pwave\ \KAp\ resonance is clear.  Complex
patterns of both constructive and destructive interference
near 1400 \MeVcc\ due to either \swave\ \KBs, \pwave\ \KBp\, 
or \dwave\ \KAd\ are also observed.
A further contribution from the \pwave\ \KCp\ state is also present,
as determined by fitting.  This is difficult to see due to smearing
of the Dalitz boundary resulting from the finite resolution in the
three-body mass.

All these resonances are well-established and are known to have
significant $\Km\pip$ partial widths.
Interference between resonances is evident in the regions of overlap.

\subsection{Asymmetry in the $\Km\pip$ System}
One of the most striking features of the Dalitz plot is the asymmetry 
in each \KAp\ band.  In any given $\Km\pip$ mass slice, a greater density
of events exists at one end of that slice than at the other.  This
asymmetry is also evident in the region closest to the \KAp\ peak itself.
This is most readily explained by interference with a $\Km\pip$ \swave\
component and clearly shows that these data can be used to
infer the structure of the \swave\ amplitude, provided an adequate
modelling of the remainder of the plot is possible.

\begin{figure}[hbt]
 \centerline{%
 \epsfig{file=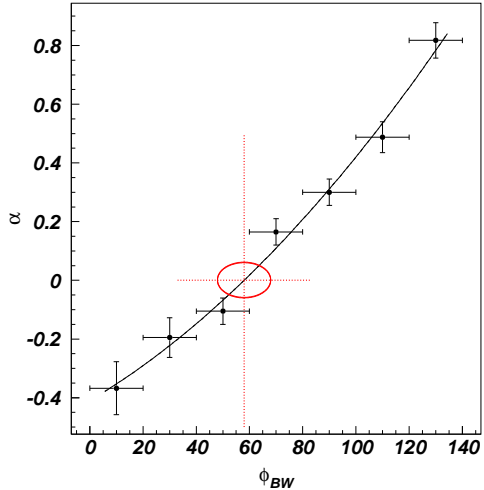,width=0.4\textwidth,angle=0}}
 \caption{%
  The asymmetry \asymm\ plotted vs. BW phase $\bwph$ for the \KAp.
  These quantities are described in the text.  \asymm\ becomes zero at
  $\bwph\sim 56$ degrees.
  \label{fig:kpi_asymb}}
\end{figure}

This asymmetry, \asymm, depends on the distribution of the helicity angle,
$\theta$, the angle between $\Km$ and $\piB$ in the $\Km\piA$ rest 
frame.  It is defined
\cite{helicity}
as
 \bea
   \alpha = {N_{\cos\theta>0} - N_{\cos\theta<0} \over
             N_{\cos\theta>0} + N_{\cos\theta<0}}
   \label{eq:asymmetry}
 \eea
where $N$ is the efficiency-corrected number of events in the
indicated regions of $\cos\theta$.
In Fig. \ref{fig:kpi_asymb}, \asymm\ is plotted as a function of the 
\KAp\ Breit-Wigner (BW)
phase $\bwph = \tan^{-1}[m_{0}\Gamma/(m_{0}^2-\SA)]$,
where the peak mass $m_{0}=896.1$~\MeVcc\ and the mass dependent width
$\Gamma=50.7$~\MeVcc\ at the peak mass.  A change in the sign of \asymm\
occurs when $\bwph\sim 56^{\circ}$, at an invariant mass below
the \KAp\ peak.  We note here that, in $K\pi$ elastic scattering
\cite{Aston:1987ir},
\asymm\ is observed to reach zero at $\bwph\approx 135.5^{\z}$, 
a mass above
the \KAp\ peak.  Evidently there is a $\sim -79^{\z}$ shift in $s$-$p$
relative phase in this $\Dp$ decay relative to that observed in $K\pi$ 
elastic scattering.  
%

%% file: method.tex
\subsection{$\Km\pip$ Partial Wave Expansion}
The Dalitz plot in Fig.~\ref{fig:dalitz_plot} is described by a
complex amplitude
Bose-symmetrized with respect to the identical pions $\piA$
and $\piB$:
 \begin{eqnarray}
   \label{eq:pwa_bose}
   \Ampl = A(\SAA,\SAB)+A(\SAB,\SAA).
 \end{eqnarray}

Considering the simplest, tree-level quark diagrams, iso-spin $I=1/2$
$\Km\pip$ systems are most likely to be produced.  The contribution of
the $\pip\pip$ amplitude to these decays is not expected to be 
significant, coming mostly from re-scattering processes.
To test this, data are taken from measurements of $\pip p\to\pip\pip n$ 
reactions
\cite{Hoogland:1977kt}
in which the phase of the $\pip\pip$ amplitude was found to vary slowly, 
assuming it to be elastic, from zero at threshold to about $-30^{\circ}$ 
at 1.45~\GeVcc, the upper range of the measurements.  No evidence for 
isospin $I=2$ resonances exists in this range.  This amplitude is added 
to those in model C in Ref.~
\cite{Aitala:2002kr}.
It is found that the $\pip\pip$ contribution is, indeed, insignificantly
small $(0.7 \pm 0.4)$\%.

The amplitude $A$ is therefore written
as the sum of $\Km\pip$ partial waves labelled by angular momentum
quantum number $L$,
 \bea
   \label{eq:pwa_expand}
   A(\SAA, \SAB) &=& \sum_{L=0}^{L_{\hbox{\tiny max}}}
      (-2pq)^L P_L(\cos\theta)\times \nonumber\\* 
      &&\FD{L}(q,\rdpl) \times C_L(\SAA),
 \eea
corresponding to production of $\Km\pip$ systems with spin $J=L$ and
parity $(-1)^L$ in these $\Dp$ decays.  
In this analysis, the sum is truncated at $L_{\hbox{\small max}}=2$ since
the \dwave\ $\KAd$, as measured in reference
\cite{Aitala:2002kr},
contributes only about 0.5\% to the decays.  This is already
small and higher partial-waves are expected to be even further
suppressed by the angular momentum barrier.  With no way to
distinguish $I=\half$ and $I={3\over 2}$ components in the $\Km\pip$
systems produced, their sum is measured in this paper.

In Eq.~(\ref{eq:pwa_expand}), $\vec p$ and $\vec q$ are momenta
for the $\Km$ and bachelor $\piB$ respectively, in the
$\Km\piA$ rest frame.  The cosine of the helicity
angle $\theta$ is then given in terms of the masses $m_{\Km}$ ($m_{\pip}$)
and energies $E_{\Km}$ ($E_{\pip}$) of the $\Km$ ($\piB$)
in the $\Km\piA$ rest frame by:
\bea
  \cos\theta &=& \hat p\cdot\hat q  \nonumber\\*
             &=& {E_{\Km}E_{\piB} -
                 \left(\SAB-m_{\Km}^2-m_{\pip}^2\right)/2
                 \over pq}.
\eea
This is the argument of the Legendre polynomial functions $P_{L}$.
$\FD{L}$ is a form factor for the parent $D$ meson which depends on $q$,
$L$ and on the $D$'s effective radius $r=\rdpl$:
\bea
  \label{eq:ffr}
  \begin{array}{lllll}
    \label{eq:ffra}
    \FD{0} &=& e^{-(r q)^2/12}  &  \hbox{scalar} &  \\
    \label{eq:ffrb}
    \FD{1} &=& \left[1 + (rq)^2\right]^{-\half} & \hbox{vector} & \\
    \label{eq:ffrc}
    \FD{2} &=& \left[9 + 3(rq)^2 + (rq)^4\right]^{-\half}
     & \hbox{tensor} &
  \end{array}
\eea
For $L>0$, these form-factors are derived for non-relativistic potential
scattering
\cite{BlattWeiss}.
For $L=0$, the Gaussian form in Eq.~(\ref{eq:ffr}), suggested by
Tornqvist 
\cite{Tornqvist:1995kr}
to be a preferred way to describe scalar systems, is used.  
This form was used also in 
Ref.~\cite{Aitala:2002kr}.

The $C_L(\SAA)$ are complex functions, and are the 
invariant-mass-dependent parts of
the respective partial waves.  They do not depend on the other Dalitz
plot variable $\SAB$ and are referred to in this paper as the $\Km\pip$ 
amplitudes.  Provided that interactions between the $\Km\piA$ system
and the bachelor $\piB$ can be neglected, the $C_L(\SAA)$ are related
to the corresponding amplitudes, 
$T_L(\SA)
$ measured in $\Km\pip$ scattering experiments, by
\bea
\label{eq:prodelastic}
   C_L(\SA) \equiv |C_L(\SA)|e^{i\phi_L(\SA)} 
   &=& {\sqrt{\SA}\over p}{\pvec_L(\SA)T_L(\SA)\over p^L\FD{L}},
\eea
where $\pvec_L(\SA)$, unknown functions, describe the $\Km\pip$ 
production in each wave in the $D$ decay process
\cite{Adler}.
These replace the
$\Km\pip$ coupling present in elastic scattering (proportional to 
the 2-body phase-space factor $\sqrt{\SA}/p$ and barrier factor $p^L$).

The principal goal of this analysis is to measure $C_0(\SA)$,
using all higher $L$ contributions to the Dalitz plot as an
``interferometer".  This requires a model for $C_1(\SA)$ and $C_2(\SA)$,
the reference \pdash\ and \ddash\ waves.

\subsection{The Reference Waves}
\label{sec:refwaves}
As in previous analyses, a Breit-Wigner isobar model is used to describe
the \pdash\ and \ddash waves.  Linear combinations of resonant propagators
$\BW_{\sst R}$, one for each of the established resonances $R$ having the
appropriate spin, and each with a
complex coupling coefficient with respect to \KAp,
$\coeff{R}=b_{\sst R}e^{i\beta_{\sst R}}$, are constructed.
Three possible $\Km\pip$ resonances are included in the \pwave, but only
one in the \dwave\ in the invariant mass range available to these decays:
%
\bea
  \label{eq:pwave}
  C_1(\SA) &= &[\BW_{\sst \KAp}(\SA) +
               \coeff{\KBp}\BW_{\sst \KBp}(\SA) +
               \nonumber\\* &&
               \coeff{\KCp}\BW_{\sst \KCp}(\SA)]
               \times \FR{L}(p,\rres)
              ,\\
  \label{eq:dwave}
  C_2(\SA) &=& \left[\coeff{\KAd}\BW_{\sst \KAd}(\SA)\right]
               \times \FR{L}(p,\rres)
               .
\eea
%
where $\FR{L}$ is a form factor for the resonances in the $\Km\pip$
system, required to ensure that the resonant amplitudes vanish for
invariant masses far above the pole masses.
It is assumed to have
the same dependence on center-of-mass momentum and angular momentum 
as the $D$ form factor $\FD{L}$, but to depend on a different effective
radius $r=\rres$.
The coefficients in Eq.~(\ref{eq:pwave}) have their origin in the 
$\Km\pip$ production process arising from $\Dp$ decays, and are therefore
treated as unknown parameters in the fits.

Each propagator is assumed to have a Breit-Wigner form defined as:
\bea
  \label{eq:bw}
    \BW_{\sst R}(\SA) = {1 \over \mR^2-\SA-i\mR\Gamma(\rres,\SA)},
\eea
where \mR\ and \GamR\ are the resonance mass and width, and:
\bea
  \label{eq:gam}
    \Gamma(\rres,\SA)=\GamR
                    \left({\mR\over\sqrt \SA}\right)
                    \left({p\over \pR}\right)^{2L+1}
                    \left[{\FR{L}(p,\rres)\over\FR{L}(\pR,\rres)}\right]^2
\eea
where $\pR$ is the value of $p$ when $\SA=\mR^2$.


\subsection{Parametrization of the \swave}
The goal is to define the \swave\ amplitude making no
assumptions about either its scalar meson composition, nor of the form
of any \swave\ $NR$ terms.  To this end, two real parameters are 
introduced
\bea
  \label{eq:ckgk}
    c_k &=& |C_0(\SAk)|~~;~~ \gamma_k = \phi_0(\SAk)
\eea
to define the amplitude $C_0(\SAk) = c_k e^{i\gamma_k}$
at each of a set of invariant mass squared values 
$\SA=\SAk~(k=1,N_s)$.
A second order spline interpolation is used to define the amplitude 
between these points $(\SAk, c_ke^{i\gamma_k})$
\cite{Spline}.
The $c_K$ and $\gamma_k$ values are treated as model-independent 
parameters, and are determined by a fit to the data.

To obtain the results in this paper, $N_s=40$ equally spaced values 
of \SAk\ are chosen.  These are indicated by the lines drawn on the
Dalitz plot in Fig.~\ref{fig:dalitz_plot}.  Other sets of values
for $\SAk$ are also used to check the stability of the results 
obtained.

\subsection{Maximum Likelihood Fit}
In this analysis, the 3-body mass $M$ is not constrained to be that
of the $\Dp$ meson.  The fits are therefore made in three dimensions 
$(M,\SAA,\SAB)$.
A normalized, log-likelihood function is defined as
 \bea
   \label{eq:ll}
   {\cal L} = \sum_{\hbox{events}}\ln\left[
                    \left(1-\sum_{i=1}^3 f_i\right) P_s +
                    \sum_{i=1}^3 f_i P^i_b\right],
 \eea
where $P_s$ and $P_b^i$ are the normalized signal and background 
PDF's, respectively.

Three backgrounds ($i=1,~2,~3$), described in Sec.~\ref{sec:dalitzplot},
are included incoherently in Eq.~(\ref{eq:ll}).  Each is considered to
constitute a fraction $f_i$ of the event sample in the selected range
$1.850<M<1.890$ \GeVcc, and to be described by the PDF:
 \bea
   \label{eq:pdfback}
   P_b^i &=& {\three_i(M)\theta_i(\SAA,\SAB)\over n_i}.
 \eea
This expression has a three-body mass profile
$\three_i(M)$ and a distribution $\theta_i(\SAA,\SAB)$, with
normalization $n_i$, over the Dalitz plot.  For the combinatorial
background, the PDF is determined by events in a band of $M$ values
above the $\Dp$ peak, while for the $D_s$ reflections it is determined 
from the simulated MC samples.
 
The signal PDF is
 \bea
   \label{eq:pdfsig}
   P_s &=& {\three_0(M)\epsilon(\SAA,\SAB)|\Ampl(\SAA,\SAB)|^2 \over
           \int d\SAA d\SAB dM~F(M)\epsilon(\SAA,\SAB)|\Ampl(\SAA,\SAB)|^2},
 \eea
in which $\three_0(M)$ describes the shape of the signal component in the 
$\Km\pip\pip$ invariant mass spectrum, parametrized as the sum of two Gaussian 
functions, and $\epsilon(\SAA,\SAB)$ is the efficiency for reconstructing 
these events.
The normalization integral extends over the entire Dalitz plot for 
each $M$ in the selected range.

\subsection{Decay Channels and Branching Fractions}
The amplitude $A(\SAA,\SAB)$ in Eq.~(\ref{eq:pwa_expand}) can be written as a
sum over the $N_{\sst ch}$ possible decay channels of the $\Dp$:
\bea
  A(\SAA,\SAB) = \sum_{k=1}^{N_{\sst ch}} A_k,
\eea
where $A_k$ is the complex amplitude for the $k^{th}$ decay mode for
decay to the $\Km\pip\pip$ final state through either a resonance,
or through the whole set of possible \swave\ and $NR$ states.
The fraction, \brat{k}, is computed for each such mode:
\bea
   \label{eq:Bk}
  \brat{k} = {\int_{\sst DP}|A_k|^2 d\SAA d\SAB \over
              \int_{\sst DP}|\sum_i A_i|^2 d\SAA d\SAB}.
\eea
This is the definition most often used in the literature on three body
decays.  It guarantees that each \brat{k}\ is positive.  Due to
interference, however, the \brat{k}\ do not necessarily sum to unity.

\subsection{Parameters, Phases and Constants}
The log-likelihood, Eq.~(\ref{eq:ll}), is defined by many parameters.
By choice, a number of these are held constant in the fits.  Parameters
for the background models $P^i_b$ and their fractions $f_i$ are 
determined by studies of data and of MC samples and are fixed.
Masses and widths for well-established \pdash\ and \dwave\ resonances
are also held constant at values listed in Table~\ref{tab:m0g0}.  These
come mostly from the Review of Particle Properties (RPP) publication
\cite{PDG}.
For the \KCp\ values appropriate for the state with known coupling
to $K\pi$ observed in $K\pi$ scattering in the LASS experiment
\cite{Aston:1987ir}
are used.  The form factor radii are fixed at
$\rdpl=5.0$~GeV$^{-1}$ and $\rres=1.6$~GeV$^{-1}$, values
determined in 
Ref.~\cite{Aitala:2002kr}
to be those providing the best isobar model description for these 
data.  Isobar coefficients $\coeff{R}$ and partial wave amplitude 
parameters $c_i$ and $\gamma_i$
are generally allowed to vary.

\begin{table}[hbt]
\caption{resonance mass \mR\ and width \GamR\ values used in the
 fits described in this paper.  With the exception of \KCp,
  parameters are as quoted in 
  Ref.~\cite{PDG}.
 \label{tab:m0g0}}
  \begin{ruledtabular}
  \begin{tabular}{ l c c }
  Resonance & \mR~(\MeVcc) & \GamR~(\MeVcc) \\
  \hline
    \KAp    &  896.1       &     50.7        \\
    \KBp    & 1414.0       &    232          \\
    \KCp    & 1677.0       &    205          \\
    \KAd    & 1432.4       &    109          \\
  \end{tabular} 
  \end{ruledtabular}
\end{table}

Phases are defined relative to the \KAp\ resonance.  In all fits
described here, the coefficient for the \KAp\ is taken to be real
and of magnitude unity, as explicit in Eq.~\ref{eq:pwave}.

Two 
sources of uncertainty in this method
result from the parametrization of the \pwave, and
from the fact that several local minima in the likelihood function
exist.  These limitations are discussed in more detail in 
Appendix~\ref{sec:ambiguities}.

%% file: swave_fit.tex
The technique described in Section \ref{sec:method} is applied to the
data shown in the Dalitz plot in Fig.~\ref{fig:dalitz_plot}.  The
\pdash\ and \dwave\ amplitudes defined as in Eqs.~(\ref{eq:pwave})
and (\ref{eq:dwave}) are chosen as reference waves.  The 40 equally
spaced values $\SAk$, indicated by lines in the figure, are chosen.
The \swave\ magnitude and phase , $c_k$ and $\gamma_k$, at each $\SAk$,
and the \pwave\ and \dwave\ couplings $\coeff{i}$
are all determined by the fit.  With all established vector and
tensor resonances with masses and widths shown in Table~\ref{tab:m0g0},
there are 86 free parameters.

It is confirmed that the contribution from \KBp\ is negligible, as
reported in
Ref.~\cite{Aitala:2002kr},
and this is dropped from further consideration.  The fit is made with the
remaining 84 free parameters.  The complex coefficients \coeff{i}\
and the fractions \brat{i}\ for each of the resonances $i$ included
in the \pdash\ and \dwave s are summarized in Table~\ref{tab:bigtab}.
\begin{table*}[hbt]
\caption{Fractions, magnitudes and phases for resonant and
 \swave\ components from four fits to decays of $\Dp$ mesons
 to $\Km\pip\pip$ described in the text.  Fit labels are ``\EIPWA"
 for the fit, described in Sec.~\ref{sec:swave}, where magnitudes
 and phases for 40 $\Km\pip$ mass slices described in the text
 are free to vary.  Systematic errors are included for this fit.
 ``Isobar" refers to the fit described in Sec.~\ref{sec:kappatest}.
 The fit labelled as ``Elastic" is described in Section
 \ref{sec:elastic}.
 \label{tab:bigtab}}
\begin{ruledtabular}
\input{bigtab.tex}
\end{ruledtabular}
\end{table*}
\cite{localmin}

\begin{figure*}[hbt]
 \centerline{%
 \epsfig{file=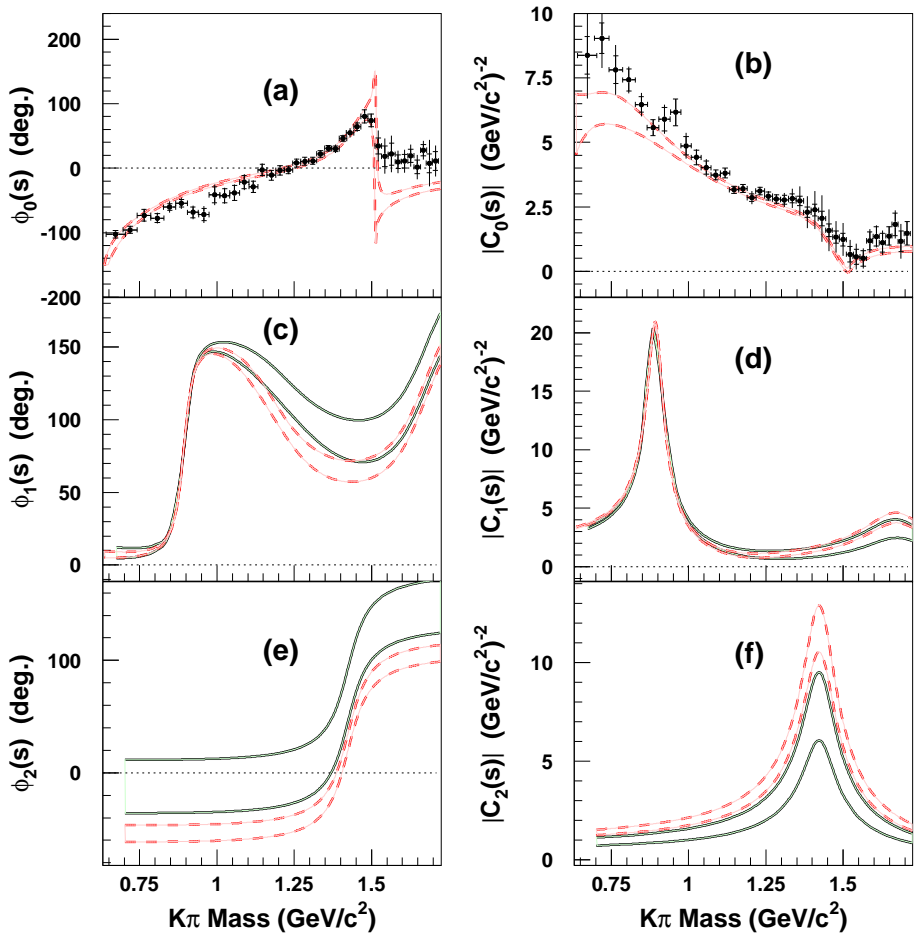,
         height=0.6\textheight,angle=0}}
 \caption{(a) Phases $\gamma_k=\phi_0(\SAk)$ and (b) magnitudes
  $c_k=|C_0(\SAk)|$ of \swave\ amplitudes for $\Km\pip$ systems from 
  $\Dp\to\Km\pip\pip$ decays with the amplitude and phase of the \KAp\ 
  as reference.  Solid circles, with error bars show the values
  obtained from the \EIPWA\ fit described in the text.  
  The effect of adding systematic uncertainties in quadrature is
  indicated by extensions on the error bars.
  The \pwave\ and \dwave\ phases are plotted in (c) and (e) and their
  magnitudes in (d) and (f), respectively.  These curves are derived
  from Eqs.~(\ref{eq:pwave}) and (\ref{eq:dwave}), respectively,
  evaluated with the parameters and error matrix resulting from the
  \EIPWA.  Curves appear as shaded areas bounded by solid line curves
  representing one standard deviation limits for these quantities.
  In all plots, the dashed curves show one standard deviation limits 
  for the predictions of the isobar model fit described in 
  Sec.~\ref{sec:kappatest}.  These curves are computed in the same way,
  using Eq.~(\ref{eq:swave}) in addition to (\ref{eq:pwave}) and 
  (\ref{eq:dwave}) with parameters and error matrix from the isobar
  model fit.
 \label{fig:solution0}}
\end{figure*}

The \swave\ phases $\gamma_k$ ($=\phi_0(\SAk)$) and magnitudes $c_k$ 
($=|C_0(\SAk)|$) resulting from the fit
are plotted, with error bars, in Figs.~\ref{fig:solution0}(a) 
and (b), respectively.  A significant phase variation is observed
over the full range of invariant mass, with the strongest variation
near the \KBs\ resonance.  The magnitude is largest just above
threshold, peaking at about 0.725~\GeVcc, above which it falls.  A
shoulder is seen at the mass of the \KBs, after which the magnitude
falls sharply to its minimum value just above 1.5~\GeVcc.

The \swave\ magnitudes $c_k$ obtained depend on the form used 
for \FD{0} in Eq.~(\ref{eq:pwa_expand}).  The products $c_k\FD{0}$,
and phases $\gamma_k$ are, however, independent of \FD{0}.  To
simplify future comparisons, values for $c_k$, $\FD{0}$
and $\gamma_k$ for each invariant mass $\SAk$ are listed, with their 
uncertainties, in Table~\ref{tab:pwa}.  In the present analysis, 
the Gaussian form
\cite{Tornqvist:1995kr}
in Eq.~(\ref{eq:ffr}) for $L=0$ has been chosen.  The values used
for \FD{0} at each $\SAk$ are also listed in Table~\ref{tab:pwa}.
%
\begin{table}[bt]
\caption{\label{tab:pwa}
 Magnitude $c$ and phase $\gamma$ of the $\Km\pip$
 \swave\ amplitude determined, at equally spaced masses, by the
 \EIPWA\ fit described in the text.  The magnitudes assume a real
 form-factor $F_{\sst 0}^{\sst D}$ for the $\Dp$ meson.  Values
 for this form-factor are given for each mass value in the table.
 Two further mass values, used in the fit as a result of the finite
 resolution in 3-body invariant mass, are not included in the table.
 They lie, respectively, at and above the kinematic limit for $\Dp$
 decay to this final state.
 Statistical and systematic uncertainties are assigned to each 
 magnitude and phase.}
\begin{ruledtabular}
\input{tab_pwa.tex}
\end{ruledtabular}
\end{table}
%

The magnitudes $|C_L(\SA)|$ and phases $\phi_L(\SA)$ for the
\pdash\ and \dwave\ amplitudes $C_L(\SA)~~(L=1,2)$ are computed
from Eqs.~(\ref{eq:pwave}) and (\ref{eq:dwave}), using parameters 
for this fit from Tables~\ref{tab:m0g0} and \ref{tab:bigtab}.
Uncertainties in these quantities are also computed, using the full 
error matrix from the fit.  Values, at each \SA, plus or minus one 
standard deviation are then plotted as solid curves, with shading 
between them, in Fig.~\ref{fig:solution0}.  The \pwave\ phase and 
magnitude are shown, respectively, in Figs.~\ref{fig:solution0}(c) 
and (d), and those for the \dwave\ in (e) and (f).

\begin{figure*}[hbt]
 \centerline{%
 \epsfig{file=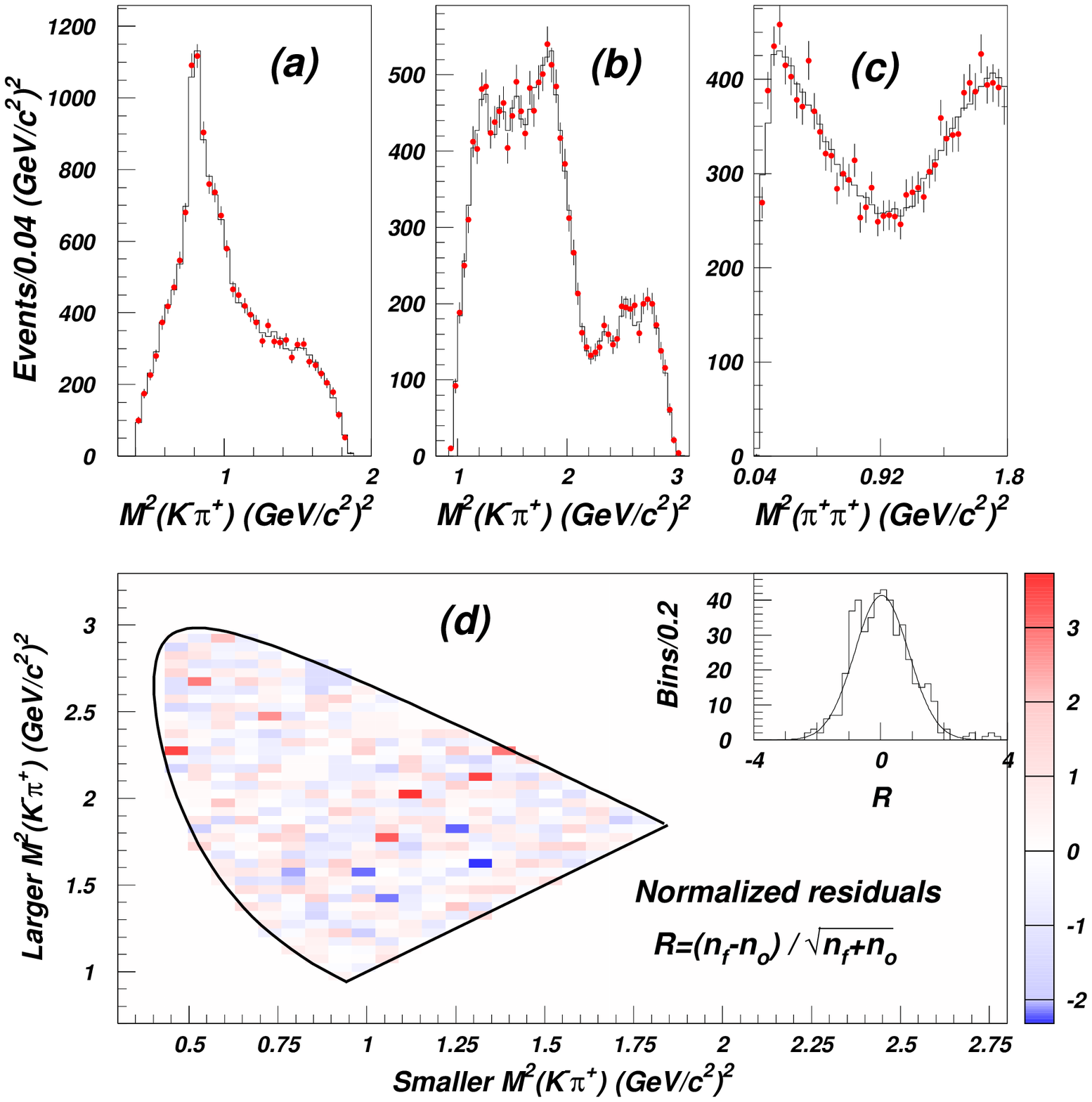,
         height=0.65\textheight,angle=0}}
 \caption{Comparison between \EIPWA\ fit and data for $\Dp\to\Km\pip\pip$
  decays.  For each event, the distributions of  (a) the smaller 
  $\Km\pip$, (b) the larger $\Km\pip$ and (c) the $\pip\pip$ squared
  invariant masses are plotted as points with error bars.  Results of 
  the fit (solid histogram) are superimposed.  (d) The Dalitz plot
  folded about the axis of symmetry resulting from the
  identity of the two $\pip$ mesons.  The quantity plotted is the
  normalized residual $(n_f-n_0)/\sqrt{n_f+n_o}$ defined in the text.  
  The histogram is the distribution of normalized residual values and 
  the curve is a Gaussian fit to this having mean
  $-0.015 \pm 0.039$ and standard deviation $0.93 \pm 0.04$.
  \label{fig:sw_pdfit}}
\end{figure*}
\begin{figure*}[hbt]
 \centerline{%
 \epsfig{file=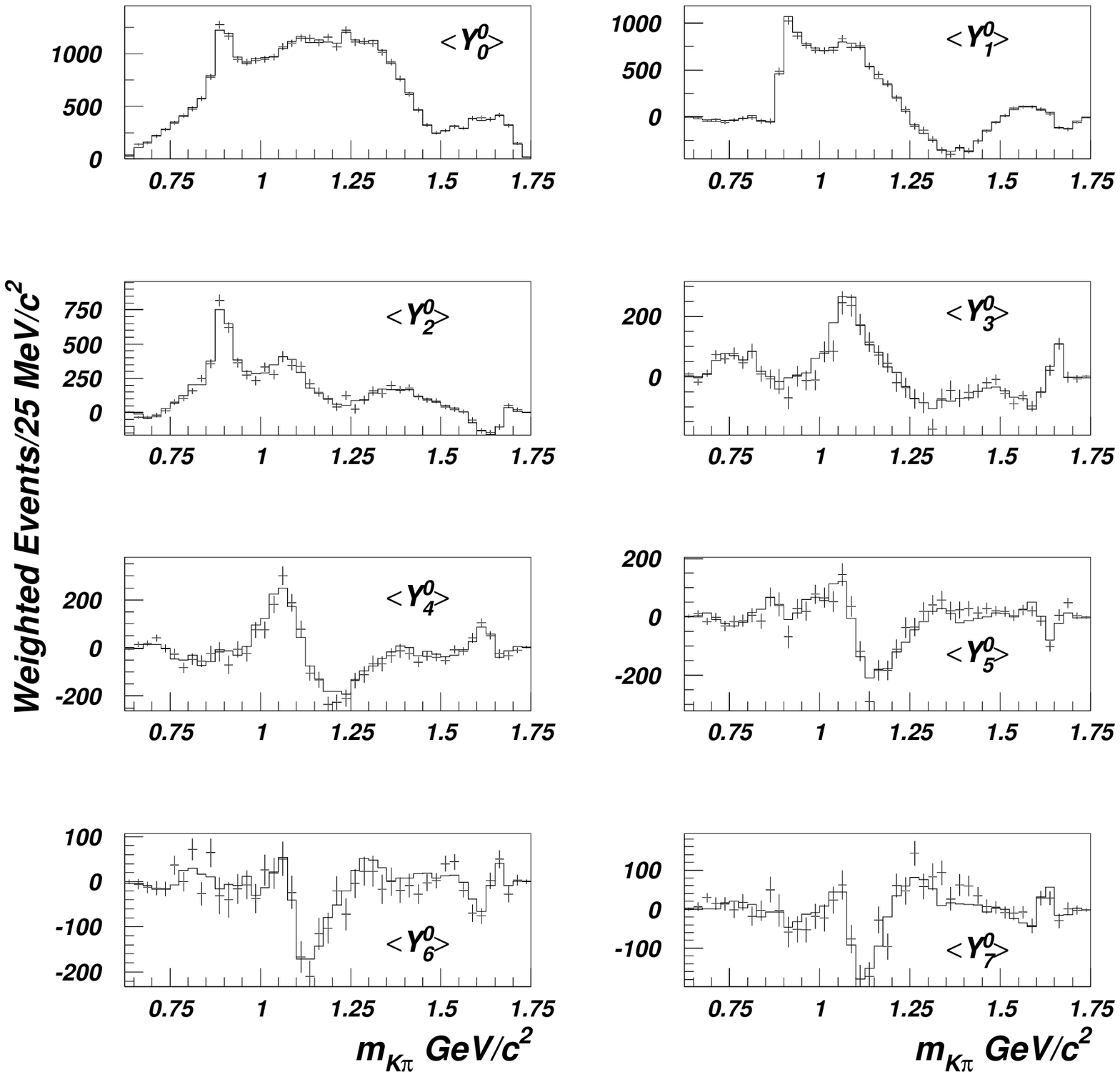,width=0.75\textwidth,angle=0}}
 \caption{Moments of the $\Km\pip$ helicity angle $\theta$ for the \EIPWA\
  fit.  In each figure, events are plotted in bins of $\Km\pip$ invariant
  mass with weights equal to $P_L(\cos\theta)$ as indicated.
  Two combinations are plotted for each $\Km\pip\pip$ candidate.  Data are
  represented as points with error bars.  Events are not weighted for their
  estimated efficiency.  The MC events, treated the same way, are used to
  show the expectation for these moments from the fit, and are
  shown as solid histograms.
 \label{fig:soln0moments}}
\end{figure*}

To compare the fit with the data, MC simulated samples of events are
produced in the three-dimensional space in which the fits are made.  
Events are generated with the distribution predicted from the signal and
background PDF's defined in Eq.~(\ref{eq:ll}).  Parameter values from
Tables~\ref{tab:bigtab}~and~\ref{tab:pwa},
and the measured event reconstruction efficiency $\epsilon(\SAA,\SAB)$,
are used in the simulation.
These events are projected onto the two-dimensional Dalitz plot, and its
one-dimensional invariant mass plots.  Data are then overlayed for
comparison.  These plots are shown in Fig.~\ref{fig:sw_pdfit}.  
As a further comparison, the distributions of the helicity angle $\theta$
in the $\Km\pip$ systems predicted by the fit are compared to the data.
Fig.~\ref{fig:soln0moments} shows moments for this
angle, $(dN/dm)\langle P_L(\cos\theta)\rangle$, uncorrected for acceptance.

Qualitatively, agreement between the fit and the data is very good.  
Quantitative comparison is made using the observed 
distribution of events on the Dalitz plot.  The plot is divided into
rectangular bins.  For each of these, the normalized residual,
$(n_f-n_o)/\sigma$, where $n_o$ is the number of events observed, $n_f$ 
is the number predicted by the fit and $\sigma=\sqrt{n_f+n_o}$ is the 
uncertainty in $n_f-n_o$, is computed.  The expected population in each
bin, $n_f$, is
computed by numerical integration of the PDF in Eq.~\ref{eq:ll}.
%
Neighboring bins are combined, where necessary, to ensure that $n_f\ge 10$.
The normalized residuals, plotted as an inset in Fig.~\ref{fig:sw_pdfit}(d),
are combined to obtain $\chidof$ where \NDF\ is the number of bins less 
the number of free parameters in the fit.
These values, and the probability for obtaining them, are tabulated with 
the optimum log-likelihood value from the fit, given in 
Table~\ref{tab:likely}.
\begin{table}[hbt]
\caption{Likelihood values for fits to the E791 $\Km\pip$ system
  from $\Dp\to\Km\pip\pip$ decays.  The fits are described in the text
  and are labelled the same way as in Table~\ref{tab:bigtab}.
  \label{tab:likely}}
\begin{ruledtabular}
\input{likely.tex}
\end{ruledtabular}
\end{table}

%% file: bigtab.tex
  \begin{tabular}{ l c c c c }
  Channel    & Fit        & Fraction             & Amplitude            & Phase                   \\[-2pt]
             &            & \brat{\phantom{.}}\% &  $b$                 & $\beta$ (degrees)               \\
  \hline 
 $\KAp\pi^+$ &\EIPWA\     &11.9$\pm  0.2\pm$ 2.0 & 1.00 (fixed)         &   0.0 (fixed)           \\[-2pt]
             &Isobar      &12.6$\pm$ 1.6         & 1.00 (fixed)         &   0.0 (fixed)           \\[-2pt]
             &Elastic     &12.8$\pm$ 2.0         & 1.00 (fixed)         &   0.0 (fixed)           \\[-2pt]
 $\KCp\pi^+$ &\EIPWA\     & 1.2$\pm  0.6\pm$ 1.2 & 1.63$\pm 0.4\pm$ 0.2 &  42.8$\pm 16.3\pm$ 4.5  \\[-2pt]
             &Isobar      & 2.1$\pm$ 0.4         & 2.18$\pm$0.2         &  28.2$\pm$ 7.2          \\[-2pt]
             &Elastic     & 5.0$\pm$ 0.8         & 3.15$\pm$0.3         &  17.1$\pm$ 7.5          \\[-2pt]
 $\KAd\pi^+$ &\EIPWA\     & 0.2$\pm  0.1\pm$ 0.1 & 4.31$\pm 1.0\pm$ 1.1 & -12.2$\pm 23.7\pm$ 16.8 \\[-2pt]
             &Isobar      & 0.5$\pm$ 0.1         & 6.50$\pm$0.7         & -54.0$\pm$ 7.4          \\[-2pt]
             &Elastic     & 0.3$\pm$ 0.1         & 4.59$\pm$0.0         & -46.9$\pm$12.3          \\[-2pt]
 Total \swave: \\
             &\EIPWA\     &78.6$\pm  1.4\pm$ 1.8 & \hss EIPWA \hss      & \hss EIPWA \hss         \\[-2pt]
             &Elastic     &79.2$\pm$ 1.1         & \hss  --   \hss      & \hss  --   \hss         \\[-2pt]
 \swave\ components:      &&                     &                      &                         \\
 $NR$        &Isobar      &16.1$\pm$ 5.3         & 0.60$\pm$0.1         &  -3.5$\pm$ 9.1          \\[-2pt]
 $\KAs\pi^+$ &Isobar      &45.6$\pm$10.7         & 1.71$\pm$0.2         & 181.3$\pm$ 8.1          \\[-2pt]
 $\KBs\pi^+$ &Isobar      &12.2$\pm$ 1.3         & 0.52$\pm$0.1         &  47.0$\pm$ 5.6          \\
  \end{tabular} 

%% file: tab_pwa.tex
  \begin{tabular}{c c rrrrr rrrrr}
$\sqrt{\SA}$   & $F_D^0(\sqrt\SA)$ & \mco{5}{c}{$c$}            & \mco{5}{c}{$\gamma$}    \\
$(\GeVcc)$     &                   & \mco{5}{c}{$(\GeVcc)^{-2}$}& \mco{5}{c}{(degrees)}   \\
  \hline
$0.672$ & $0.26$ & $8.37$ &$\pm$& $0.73$ &$\pm$& $0.62$ & $-102$ &$\pm$& $ 5$ &$\pm$& $3$ \\[-3pt]
$0.719$ & $0.27$ & $9.04$ &$\pm$& $0.59$ &$\pm$& $0.89$ & $ -96$ &$\pm$& $ 5$ &$\pm$& $3$ \\[-3pt]
$0.764$ & $0.29$ & $7.82$ &$\pm$& $0.54$ &$\pm$& $0.89$ & $ -73$ &$\pm$& $ 9$ &$\pm$& $4$ \\[-3pt]
$0.807$ & $0.31$ & $7.42$ &$\pm$& $0.43$ &$\pm$& $0.57$ & $ -77$ &$\pm$& $ 7$ &$\pm$& $5$ \\[-3pt]
$0.847$ & $0.33$ & $6.47$ &$\pm$& $0.30$ &$\pm$& $0.46$ & $ -60$ &$\pm$& $ 5$ &$\pm$& $6$ \\[-3pt]
$0.885$ & $0.34$ & $5.57$ &$\pm$& $0.31$ &$\pm$& $0.07$ & $ -54$ &$\pm$& $ 6$ &$\pm$& $5$ \\[-3pt]
$0.922$ & $0.36$ & $5.90$ &$\pm$& $0.46$ &$\pm$& $0.09$ & $ -68$ &$\pm$& $ 8$ &$\pm$& $7$ \\[-3pt]
$0.958$ & $0.38$ & $6.17$ &$\pm$& $0.52$ &$\pm$& $0.01$ & $ -72$ &$\pm$& $10$ &$\pm$& $9$ \\[-3pt]
$0.992$ & $0.40$ & $4.87$ &$\pm$& $0.35$ &$\pm$& $0.19$ & $ -41$ &$\pm$& $12$ &$\pm$& $10$\\[-3pt]
$1.025$ & $0.42$ & $4.42$ &$\pm$& $0.28$ &$\pm$& $0.09$ & $ -43$ &$\pm$& $11$ &$\pm$& $5$ \\[-3pt]
$1.057$ & $0.44$ & $4.02$ &$\pm$& $0.26$ &$\pm$& $0.01$ & $ -38$ &$\pm$& $12$ &$\pm$& $5$ \\[-3pt]
$1.088$ & $0.46$ & $3.74$ &$\pm$& $0.19$ &$\pm$& $0.11$ & $ -22$ &$\pm$& $10$ &$\pm$& $4$ \\[-3pt]
$1.118$ & $0.49$ & $3.81$ &$\pm$& $0.19$ &$\pm$& $0.13$ & $ -29$ &$\pm$& $ 9$ &$\pm$& $4$ \\[-3pt]
$1.147$ & $0.51$ & $3.16$ &$\pm$& $0.14$ &$\pm$& $0.13$ & $  -3$ &$\pm$& $ 9$ &$\pm$& $4$ \\[-3pt]
$1.176$ & $0.53$ & $3.21$ &$\pm$& $0.15$ &$\pm$& $0.13$ & $ -11$ &$\pm$& $ 7$ &$\pm$& $3$ \\[-3pt]
$1.204$ & $0.55$ & $2.86$ &$\pm$& $0.14$ &$\pm$& $0.32$ & $  -3$ &$\pm$& $ 7$ &$\pm$& $3$ \\[-3pt]
$1.231$ & $0.58$ & $3.11$ &$\pm$& $0.15$ &$\pm$& $0.13$ & $  -3$ &$\pm$& $ 6$ &$\pm$& $2$ \\[-3pt]
$1.258$ & $0.60$ & $2.92$ &$\pm$& $0.15$ &$\pm$& $0.24$ & $   8$ &$\pm$& $ 6$ &$\pm$& $3$ \\[-3pt]
$1.284$ & $0.62$ & $2.80$ &$\pm$& $0.16$ &$\pm$& $0.18$ & $  11$ &$\pm$& $ 6$ &$\pm$& $2$ \\[-3pt]
$1.310$ & $0.65$ & $2.77$ &$\pm$& $0.17$ &$\pm$& $0.12$ & $  11$ &$\pm$& $ 5$ &$\pm$& $2$ \\[-3pt]
$1.335$ & $0.67$ & $2.83$ &$\pm$& $0.17$ &$\pm$& $0.20$ & $  22$ &$\pm$& $ 5$ &$\pm$& $2$ \\[-3pt]
$1.360$ & $0.69$ & $2.73$ &$\pm$& $0.19$ &$\pm$& $0.31$ & $  31$ &$\pm$& $ 4$ &$\pm$& $2$ \\[-3pt]
$1.384$ & $0.71$ & $2.29$ &$\pm$& $0.20$ &$\pm$& $0.25$ & $  30$ &$\pm$& $ 5$ &$\pm$& $2$ \\[-3pt]
$1.408$ & $0.74$ & $2.38$ &$\pm$& $0.23$ &$\pm$& $0.01$ & $  46$ &$\pm$& $ 4$ &$\pm$& $2$ \\[-3pt]
$1.431$ & $0.76$ & $2.05$ &$\pm$& $0.28$ &$\pm$& $0.08$ & $  55$ &$\pm$& $ 4$ &$\pm$& $2$ \\[-3pt]
$1.454$ & $0.78$ & $1.59$ &$\pm$& $0.25$ &$\pm$& $0.07$ & $  64$ &$\pm$& $ 6$ &$\pm$& $4$ \\[-3pt]
$1.477$ & $0.80$ & $1.33$ &$\pm$& $0.24$ &$\pm$& $0.01$ & $  80$ &$\pm$& $10$ &$\pm$& $4$ \\[-3pt]
$1.499$ & $0.82$ & $1.23$ &$\pm$& $0.24$ &$\pm$& $0.01$ & $  74$ &$\pm$& $10$ &$\pm$& $4$ \\[-3pt]
$1.522$ & $0.84$ & $0.66$ &$\pm$& $0.30$ &$\pm$& $0.27$ & $  34$ &$\pm$& $13$ &$\pm$& $21$\\[-3pt]
$1.543$ & $0.86$ & $0.57$ &$\pm$& $0.29$ &$\pm$& $0.11$ & $  18$ &$\pm$& $16$ &$\pm$& $22$\\[-3pt]
$1.565$ & $0.88$ & $0.50$ &$\pm$& $0.30$ &$\pm$& $0.01$ & $  22$ &$\pm$& $17$ &$\pm$& $23$\\[-3pt]
$1.586$ & $0.90$ & $1.18$ &$\pm$& $0.35$ &$\pm$& $0.01$ & $  10$ &$\pm$& $10$ &$\pm$& $14$\\[-3pt]
$1.607$ & $0.92$ & $1.35$ &$\pm$& $0.37$ &$\pm$& $0.18$ & $  11$ &$\pm$& $10$ &$\pm$& $14$\\[-3pt]
$1.627$ & $0.93$ & $1.11$ &$\pm$& $0.37$ &$\pm$& $0.14$ & $  19$ &$\pm$& $10$ &$\pm$& $14$\\[-3pt]
$1.648$ & $0.95$ & $1.37$ &$\pm$& $0.35$ &$\pm$& $0.01$ & $   2$ &$\pm$& $10$ &$\pm$& $14$\\[-3pt]
$1.668$ & $0.96$ & $1.82$ &$\pm$& $0.43$ &$\pm$& $0.01$ & $  28$ &$\pm$& $ 8$ &$\pm$& $12$\\[-3pt]
$1.687$ & $0.98$ & $1.16$ &$\pm$& $0.40$ &$\pm$& $0.84$ & $   8$ &$\pm$& $14$ &$\pm$& $34$\\[-3pt]
$1.707$ & $0.99$ & $1.47$ &$\pm$& $0.46$ &$\pm$& $0.01$ & $  11$ &$\pm$& $14$ &$\pm$& $21$\\[-3pt]
  \end{tabular} 

%% file: likely.tex
\begin{tabular}{ l c c c c c c }
Model & $\ln({\cal L})$ & Number of  & $NDF$ & $\chi^2/NDF$ & Probability\\
      &                 & Variables  &       &              &            \\
\hline
\EIPWA\ fit & 36121  &    86      &  277  &    1.00 &  47.8\%  \\
Isobar      & 36072  &    16      &  412  &    1.08 &  13.2\%  \\
Elastic     & 36092  &    44      &  300  &    0.99 &  54.9\%  \\
Unitary     & 36004  &    44      &  195  &    2.68 & $\sim 0$ \\
\end{tabular}

%% file: kappatest.tex
It is interesting to compare the results from the \EIPWA\ with those 
reported in
Ref.~\cite{Aitala:2002kr}
which came from a Breit-Wigner isobar model fit.
In this fit, the \swave\ was modelled
as a sum of isobars with Breit-Wigner propagators for the \KBs\ resonance,
and another \KAs\ state.  An ``NR" term, 
defined as a constant everywhere on the Dalitz plot, was also included
in the \swave\
\bea
  \label{eq:swave}
  C_1(\SA) &=& \left[NR +
               \coeff{\KAs}\BW_{\sst \KAs}(\SA) +
               \coeff{\KBs}\BW_{\sst \KBs}(\SA)\right] \nonumber\\*
           & & \times \FR{L}(p,\rres).
\eea
The \pdash\ and \dwave s were defined as in Eqs.~(\ref{eq:pwave})
and (\ref{eq:dwave}).

For purposes of comparison, this fit is made again, exactly
as before, except that the resonance parameters indicated in 
Table~\ref{tab:m0g0} are used to replace those from
Ref.~\cite{Aitala:2002kr}.
Both the \KAs\ and \KBs\ isobars included in the \swave\ in 
Eq.~(\ref{eq:swave}) have masses and widths that are allowed to vary.
The phase convention is defined, as before, by Eq.~(\ref{eq:pwave}).  
As found in
Ref.~\cite{Aitala:2002kr},
the amplitude and fraction for \KBp\ are negligibly small.  This
resonance is, therefore, also omitted from this fit which is
labelled the ``isobar fit".  The couplings and
fractions obtained are summarized in Table~\ref{tab:bigtab}.
It is seen that the $NR$ term contributes modestly to the decays in 
this model.  Its presence is, however, important as it interferes
with the \KAs, destructively at $\Km\pip$ threshold, not at all at
780~\MeVcc, and constructively at higher mass.  Without the $NR$
term, the \swave\ form does not fit the data well.
All these results, including
Breit-Wigner masses and widths obtained for
the \swave\ states,
agree well, within uncertainties, with those
in Ref.~\cite{Aitala:2002kr}.

Amplitudes from this fit are plotted in 
Figs.~\ref{fig:solution0}(a)-(f) where they may be compared
with the \EIPWA\ results.  As for the \EIPWA, 
Eqs.~(\ref{eq:pwave}) and (\ref{eq:dwave}) are used, this time
with parameters for the isobar fit in Table~\ref{tab:bigtab}
to compute the magnitudes $|C_L(\SA)|$ and phases $\phi_L(\SA)$ 
for the \pdash\ and \dwave, respectively, for $L=0$ and $1$.  
Eq.~(\ref{eq:swave}) is used in the same way to compute the 
\swave\ amplitude.  Uncertainties in magnitudes and in phases are 
computed using the full error matrix from the isobar fit and 
values at each \SA, plus or minus one standard deviation, are 
plotted as dashed curves, with shading between them, in the 
appropriate entries in Fig.~\ref{fig:solution0}.

The isobar fit constrains the \swave\ magnitude and phase to assume
the functional forms specified in Eq.~(\ref{eq:swave}) while the 
\EIPWA\ allows them complete freedom.  
Because of the additional degrees of freedom, the latter is therefore 
able to achieve a better description of the data by
a combination of shifts in the \pdash\ and \dwave\ parameters, and
in the $(c_k, \gamma_k)$ values for the \swave.  The results presented
in Figs.~\ref{fig:solution0}(a)-(f), and in Table~\ref{tab:bigtab},
illustrate this.  Small differences between the fits in parameters
for \KCp\ and \KAd\ result in relatively large shifts in the curves
shown in Figs.~\ref{fig:solution0}(c)-(f).  These changes propagate
to the \swave.

The shapes predicted by both fits for the \swave\ phase and magnitude,
are shown in Figs.~\ref{fig:solution0}(a) and (b).
Some differences are seen in magnitudes from
$\Km\pip$ threshold up to about 900~\MeVcc, and in both phase and 
magnitude above the \KBs\ resonance.  These effects are correlated 
with one another and with the differences in the \pdash\ and \dwave s
noted above.  Similar effects are observed in tests made on
a large number of MC samples, with sizes similar to that
of the data.  Approximately 15\% of these samples, generated with
the distribution predicted by the isobar fit, give \EIPWA\ results
with similar shifts in \pdash\ and \dwave\ parameters, and in the 
associated differences in \swave\ observed in the data.
The MC tests are discussed in Appendix~\ref{sec:ambiguities}.

The significance of any differences between amplitudes obtained in 
the two fits is evaluated by comparing their abilities to describe 
distributions of kinematic quantities observed in the data.  Plots 
similar to those in Figs.~\ref{fig:sw_pdfit} and 
\ref{fig:soln0moments} are made showing similar, excellent agreement 
between fit and data.  
Using the method described in the 
previous Section, the distribution observed on the Dalitz plot is 
compared, quantitatively, with that described by the isobar fit 
results.  A value for $\chidof=1.08$ is obtained, and can be 
compared with $\chidof=1.00$ for the \EIPWA.  These results are 
included in Table~\ref{tab:likely}.  
Differences between the two fits in predicted populations of bins
in the Dalitz plot are all less than their statistical uncertainties.  
It is evident that both \EIPWA\ and isobar 
fits are good and that no statistically significant distinction 
between these two descriptions of the data can be drawn with a 
sample of this size.

%% file: elastic.tex
It is interesting to compare the amplitudes $C_L(\SA)$ defined in 
Sec.~\ref{sec:method} and measured in Sec.~\ref{sec:swave} with 
those from $\Km\pip$ scattering, $T_L(\SA)$.  The relationship between
$C_L$ and $T_L$ is given by Eq.~(\ref{eq:prodelastic}).  
If the $\Km\piA$ systems produced in $\Dp\to\Km\piA\piB$
decays do not interact with the bachelor $\piB$, then
the factor $\pvec_L(\SA)$ describes the production of $\Km\pip$ as
a function of \SA\ from these decays.  Also, under the same assumptions,
the Watson theorem
\cite{watson:1952ji} 
requires that, in the \SA\ range where $\Km\pip$ scattering is purely
elastic, $\pvec_L(\SA)$ for each partial wave labelled by $L$ and 
by iso-spin $I$, should carry no $\SA$-dependent phase.  In other words,
$\phi_L$, the phase of $C_L(\SA)$ for each partial wave, should differ, 
at most, by a constant relative to that of the corresponding elastic
scattering amplitude $T_L(\SA)$.
The magnitudes $|C_L(\SA)|$ and $|T_L(\SA)|$ could differ, however, 
due to any \SA-dependence of the production rate of $\Km\pip$
systems in $\Dp$ decay.

The validity of the Watson theorem therefore relies on the assumption
that no final state scattering between $(\Km\piA)$ and $\piB$ occurs.
This assumption, for decays such as those studied
here in which the final state consists of strongly interacting
particles, has often been assumed to hold.  However, it has never been
objectively tested.  The \EIPWA\ results from the present data provide,
therefore, an interesting opportunity to make such a test, and also
to examine the form for the production factor $\pvec_L(\SA)$.

\subsection{$\Km\pip$ Scattering}
In the \swave, below $K\eta^{\prime}$ threshold at 1.454~\GeVcc, $\Km\pip$ 
scattering in both the isospin $I=\half$ and $I={3\over 2}$ $K\pi$ 
amplitudes is predominantly elastic.  Scattering into $K\eta$ at a 
lower threshold is strongly suppressed by the
$SU(3)_{\hbox{\tiny flavor}}$ coupling to this channel.  This has been
confirmed by the LASS collaboration in energy independent measurements of
partial wave amplitudes for $\Km\pip$ scattering through, and beyond this
range
\cite{Aston:1987ir}.
$I=\half$ components of \sdash, \pdash\ and \dwave s were extracted from 
the total using measurements of the $I={3\over 2}$ scattering from
$\Kp\to\Kp\pip n$ data
Ref.~\cite{Estabrooks:1977xe}.
Scattering in higher angular momentum waves can become inelastic at the
lower $K\pi\pi$ threshold.  It was observed, however, that, in the LASS data,
\pwave\ scattering remained elastic up to approximately 1050~\MeVcc.  For
the \dwave, no significant elastic scattering was observed.

In the elastic region, the $I=\half$ component of the \swave\ $\Km\pip$ 
amplitude was fit, by the LASS collaboration, to a unitary form
\bea
  \label{eq:unitarity}
  T_0(\SA) &=& \sin[\gamma(\SA)-\gammaz] e^{i[\gamma(\SA)-\gammaz]},
\eea
where the phase $\gamma = \gamma_{\sst R} + \gamma_B + \gammaz$ is made up
from three contributions:
\bea
  \label{eq:lassdelta}
  \begin{array}{lcl}
    \cot\gamma_{\sst B} &=& {1\over pa} + \half bp, \\
    \cot\gamma_{\sst R} &=&
     {m_{\sst R}^2-\SA\over m_{\sst R}\Gamma(\rres,\SA)}, \\
    \gammaz             &=& 0~~\hbox{(arbitrary offset)}.
  \end{array}
\eea
The first is a non-resonant contribution defined by a scattering length 
$a$ and an effective range $b$.  The second contribution $\gamma_{\sst R}$
has parameters $m_{\sst R}$ and $\Gamma_{\sst R}$, the mass and width of 
the \KBs\ resonance.  In the LASS analysis, the arbitrary phase 
$\gammaz$\ was set to zero.
The $I=\half$ \pwave\ and \dwave\ amplitudes measured by LASS were found 
to be significant in this invariant mass range.

\subsection{Test of the Watson Theorem}
\begin{figure*}[hbt]
 \centerline{%
 \epsfig{file=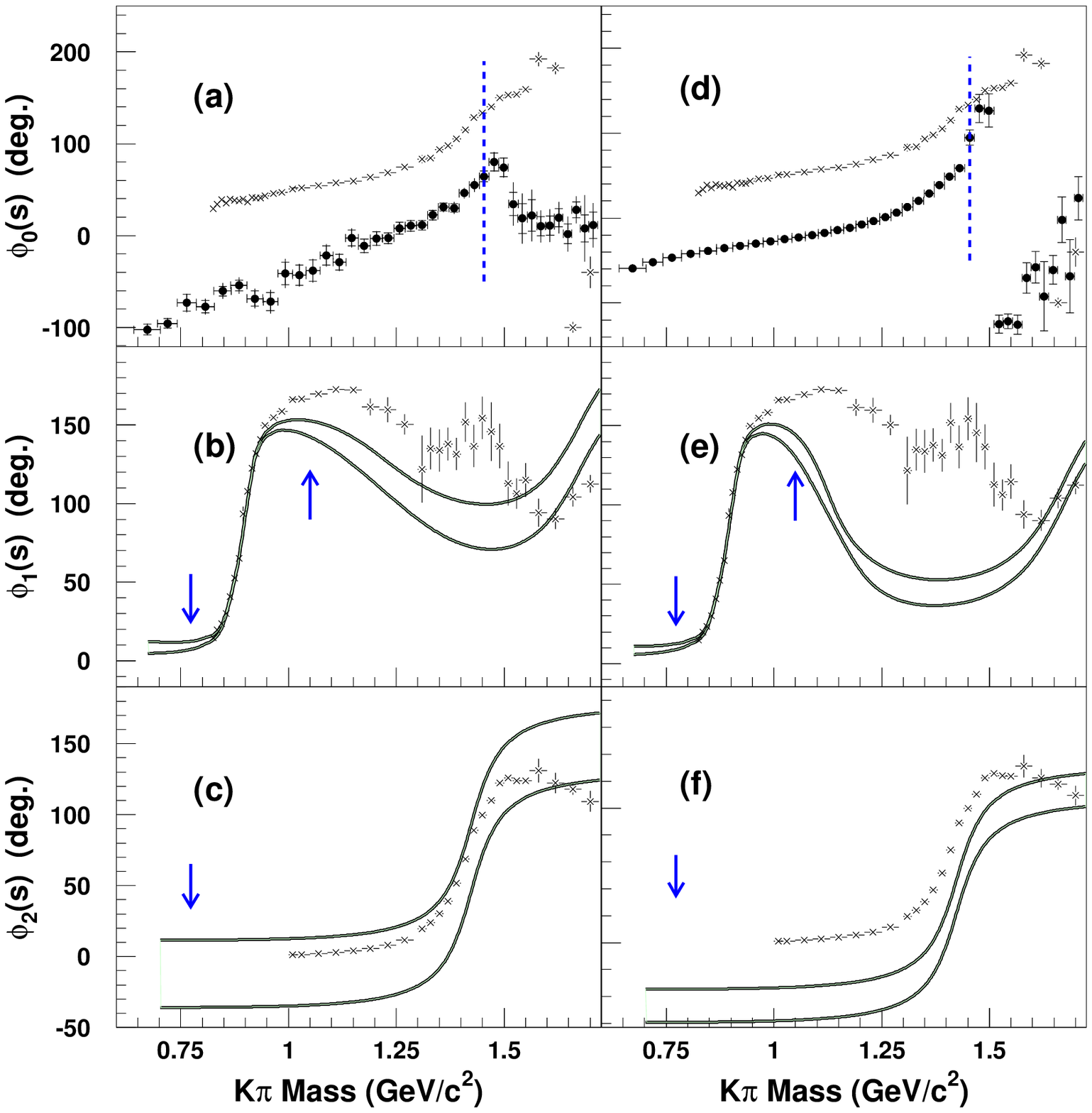,
              width=0.70\textwidth,angle=0}
 }
 \caption{Plots show comparisons between results obtained from the 
  \EIPWA, described in Sec.~\ref{sec:swave}, for phases of the
  (a) \swave, (b) \pwave\ and (c) \dwave\ $\Km\pip$ scattering 
  amplitudes with measurements made by the LASS collaboration
  \cite{Aston:1987ir}.
  Solid circles in (a) are the phases $\gamma_k$ obtained at
  each invariant mass \SAk\ from the \EIPWA.
  The vertical dashed line at the $K\eta'$ threshold (1454~\MeVcc)
  indicates the upper limit of
  the range where $\Km\pip$ scattering is predominantly elastic in
  the \swave.
  Solid curves in (b) and (c) enclose the zones 
  within one standard deviation of the \EIPWA\ \pdash\ and \dwave s,
  computed as described in Sec.~\ref{sec:swave}.
  Arrows indicate the position of the $K\pi\pi$ threshold, at which
  $\Km\pip$ scattering in the \pdash\ and \dwave s can become inelastic.
  In (b) a further arrow at $1050$~\MeVcc\ indicates the approximate
  invariant mass at which the \pwave\ data from Ref.~\cite{Aston:1987ir} 
  were observed to become inelastic.
  In (d)-(f) the results from the ``elastic fit" are shown in place of
  those from the \EIPWA.
  In all plots $I=\half$ measurements from the LASS experiment
  of phases for $\Km\pip$ scattering in these waves are shown as
  $\times$'s with error bars indicating statistical uncertainties.
 \label{fig:spd_lass}}
\end{figure*}

In Figs.~\ref{fig:spd_lass}(a)-(c), direct comparisons are made,
respectively, between the
\sdash, \pdash\ and \dwave\ phases determined by the \EIPWA\ fit to data
from this experiment and the $I=\half$ data from LASS.  The \swave\ phase
measurements, and the curves for the \pdash\ and \dwave s resulting 
from the \EIPWA, previously shown in Fig.~\ref{fig:solution0}, are
plotted, respectively, in Figs.~\ref{fig:spd_lass}(a), (b) and (c).
The LASS measurements are superimposed, as $\times$'s with error
bars, in the appropriate places in the figure.

An obvious feature in the comparison is the overall shift in phase of
the \swave\ in these data relative to that in the LASS measurements.
This feature is also evident from the examination of the $\Km\pip$
asymmetry in the Dalitz plot reported in Sec.~\ref{sec:dalitzplot}.
Another feature of the \swave\ comparison is that, for invariant
masses near $K\pi$ threshold, the phases for the two sets of data 
show a somewhat different dependence on \SA.

The \pwave s also differ in the mass range from the \KAp\ peak, through
the region where LASS observed \KBp\ scattering to become inelastic, at
approximately 1050~\MeVcc.  This difference may arise, in part, from the
parametrization of this wave given in Eq.~\ref{eq:pwave}.  With
more than one resonance described by Breit-Wigner propagators, this
may not be unitary.  The phase measured in the \dwave\ in this
experiment agrees well with that measured by LASS.  However, as
verified by the LASS data, the scattering in this wave is no longer 
elastic beyond $\KAp\pip$ threshold.

The observed shift in \swave\ phase and difference in slope, and the
difference in \pwave\ phase behaviour evidenced in
Figs.~\ref{fig:spd_lass}(a)-(c) do not conform to the precise expectations
of the Watson theorem.

\subsection{Fit with LASS Model for \swave\ Phase}
Some of the discrepancies noted above could arise from the modelling
of the \pwave.  A different model could result in a different dependence
on \SA\ of the \swave\ measured here.
To judge the significance of the observed discrepancies, therefore,
a fit is made to the data in which the \swave\ phase is constrained
to precisely follow the LASS parametrization in 
Eqs.~(\ref{eq:unitarity}-\ref{eq:lassdelta}) for invariant masses below
$K\eta'$ threshold.  The mass and width of the \KBs\ and the parameters 
$a$ and $b$ are required to assume the values obtained by LASS.  However,
the overall phase $\gammaz$, all phases above $K\eta'$ threshold, all 
magnitudes throughout the entire range of \SA, and the complex couplings 
for \pdash\ and \dwave s are determined by the fit.

This is labelled as the ``elastic fit".  A value 
$\gammaz=(-74.4\pm 1.8\pm 1.0)^{\circ}$ is obtained.  The isobar 
couplings and resonance fractions obtained are listed in 
Table~\ref{tab:bigtab}.  The \KCp\ resonance has a more significant 
contribution to this fit than in the \EIPWA.

The phases obtained for the three partial waves from the elastic fit
are compared, in Figs.~\ref{fig:spd_lass}(d)-(f), with those measured in
the LASS experiment.  The comparison is shown in the same way as in
Figs.~\ref{fig:spd_lass}(a)-(c) for the \EIPWA\ fit.
The shape of the \swave\ phase is, as required in this fit, in perfect
agreement with the LASS results.  The large offset in overall phase,
$\gammaz$ does, however persist.  Additionally, both the \pdash\ and \dwave\ 
phases now show larger differences than before.  The \dwave\ phase shifts
by $\sim 50^{\z}$, and the \pwave\ phase shows significant differences 
in the region between the \KAp\ peak and the effective limit of elastic
scattering at $\sim 1050$~\MeVcc.
%

This fit provides another excellent description of the data,
with $\chidof=0.99$ and probability 55\%, as recorded in 
Table~\ref{tab:likely}. This is comparable with both the isobar and 
\EIPWA\ fits.  However, if these observations are predominantly of 
production of $I=\half$ $\Km\pip$ systems, the phase variation 
required by the Watson theorem, is not observed in these data.

%
%

\subsection{Production Rate for the $\Km\pip$ System}

For purely elastic scattering, the $T_L$ amplitudes are required to be
unitary, as given by Eq.~(\ref{eq:unitarity}).  Introducing this into
Eq.~(\ref{eq:prodelastic}) leads to
\bea
  \label{eq:swelastic}
    C_L(\SA)
    &=& \left({\sqrt{\SA}\over p}\right)
        \left({\pvec_L(\SA)\over p^L\FD{L}}\right)
        \times\nonumber\\*
    & & \sin[\gamma(\SA)-\gammaz] e^{i[\gamma(\SA)-\gammaz]}. \\
  \hbox{For $L=0$} \nonumber\\*
  \label{eq:production}
    |\pvec_0(\SA)| &=& \left({p\over\sqrt{\SA}}\right)
    \left({\FD{0}|C_0(\SA)|\over\sin[\gamma(\SA)-\gammaz]}\right)
\eea
Structure in the \SA -dependence of the \swave\ magnitude,
$C_0(\SA)$ can thus come either from the phase
$\gamma(\SA)$, from $\pvec_0(\SA)$, or from both.  It is of
interest to study these possibilities, and to see if the data
can be described by a unitary amplitude, in which 
$\pvec_0(\SA)$ would be independent of \SA.

%
The data from the \EIPWA\ are examined to see if the \swave\ can be
described by a unitary amplitude, such as that given in 
Eq.~(\ref{eq:swelastic}).  Setting $L=0$ and $\pvec_0(\SA)=P$ (a constant),
a value for $\gammaz$ is determined by minimizing the quantity
\bea
  \label{eq:find_gammaz}
    \chi^2 = \sum_{k=1}^{N_{K\eta^{\prime}}}
             \left({|\pvec_0(\SAk)|-P\over
                    \sigma(\pvec_0)}\right)^2,
\eea
where $|\pvec_0(\SAk)|$ are computed from Eq.~(\ref{eq:production}),
for the values of \swave\ amplitude, $C_0(\SAk)=c_k e^{i\gamma_k}$, 
determined by the \EIPWA\ fit, and $\sigma(\pvec_0)$ are the
associated uncertainties.  The summation in Eq.~(\ref{eq:find_gammaz})
is made only for the $N_{K\eta^{\prime}}$ values of \SAk\ up to the
$\Km\eta^{\prime}$ threshold.  
The value $\gammaz = (-123.3\pm 3.9)^{\circ}$ is obtained, 
with $P = 0.74\pm 0.01~(\GeVcc)^{-2}$.  Fig.~\ref{fig:solution0}(a)
shows that this value for $\gammaz$ is approximately equal to the 
measured \swave\ phase at $\Km\pip$ threshold, consistent with the
physical meaning of this parameter in the formulation in 
Eq.~(\ref{eq:swelastic}).

\begin{figure}[hbt]
  \centerline{%
  \epsfig{file=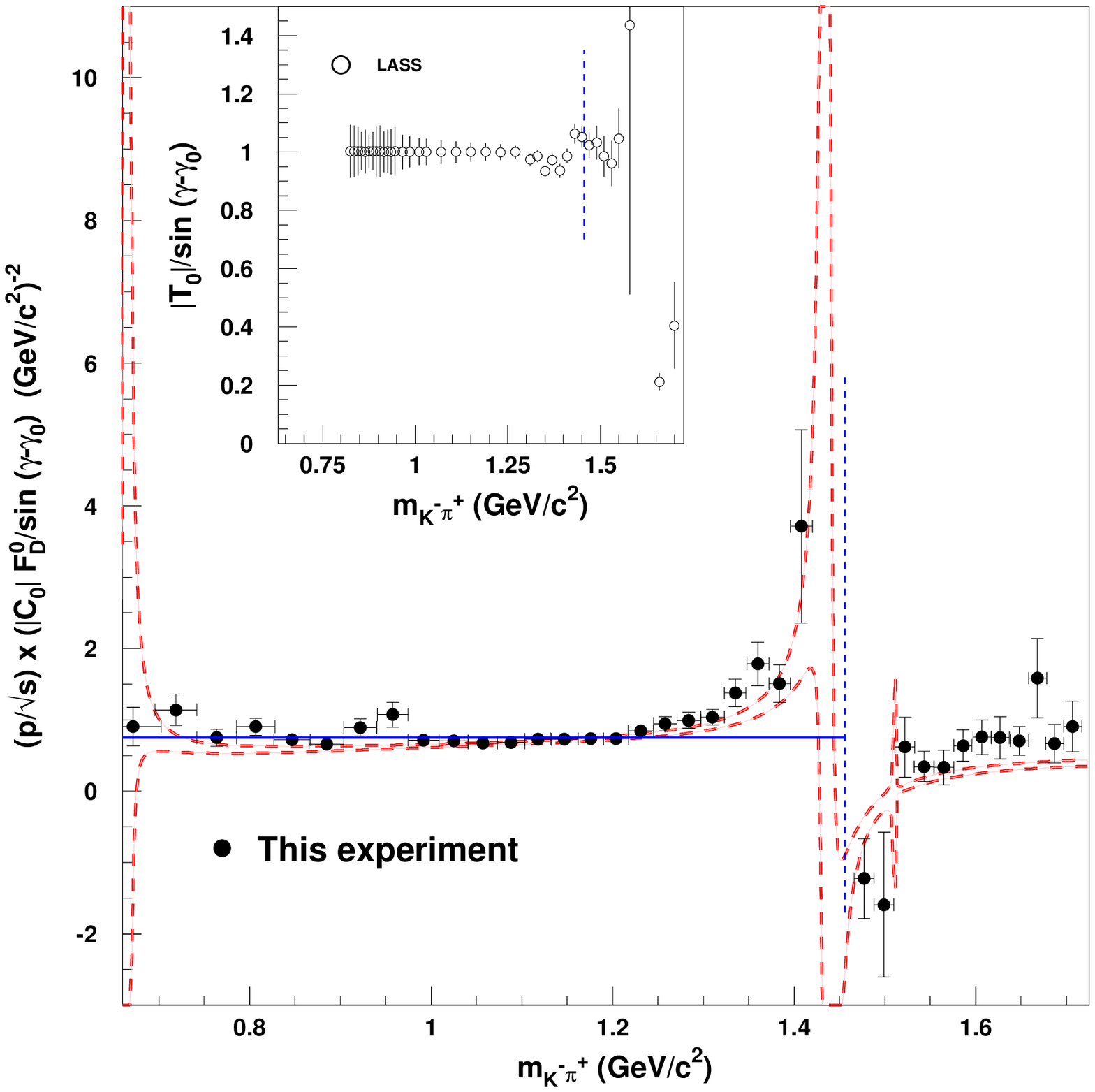,
               width=0.5\textwidth,angle=0}}
  \caption{The quantities 
   $p/\sqrt(\SAk)\times|C_0(\SAk)|\times\FD{0}/\sin(\gamma_k-\gammaz)$
   plotted as solid circles for each point obtained for the \swave\
   amplitude in the \EIPWA\ fit described in Sec.~\ref{sec:swave}.
   Three points between 1400 and 1450~\MeVcc\ are omitted from
   the plot as their values for $\sin(\gamma_k-\gammaz)$ are
   very small.  Their values are either off-scale or their
   errors extremely large.
   The region between the dashed lines shows the one standard 
   deviation limits of this quantity for the \swave\ amplitude 
   obtained from the isobar fit.  The inset shows, as small open 
   circles, the quantities
   $|T_0(\SA)|/\sin(\gamma_{\sst B}+\gamma_{\sst R})$
   taken from the LASS experiment.  
   The $K\eta^{\prime}$ threshold is indicated by dashed,
   vertical lines in both plots.
  \label{fig:smag_pvec}}
 \end{figure}

Inserting this value for $\gammaz$ into Eq.~(\ref{eq:production}),
the quantities $|\pvec_0{\SAk}|$ are plotted in 
Fig.~\ref{fig:smag_pvec}.  The solid, horizontal
line indicates the value for $P$ obtained from the fit.  The 
points are seen to lie close to this line, showing very little 
dependence on \SA\ in the invariant mass range from $\Km\pip$ 
threshold up to about 1.25~\GeVcc.  From 1.25~to~1.5~\GeVcc, 
strong variation is observed.  In this region, as seen in 
Fig.~\ref{fig:solution0}(a), the value of $\sin(\gamma-\gammaz)$, 
which appears in the denominator of Eq.~(\ref{eq:production}), is 
approximately zero.

Also shown in Fig.~\ref{fig:smag_pvec} is $|\pvec_0(\SA)|$ for
the isobar fit.  The region between dashed lines corresponds to
the one standard deviation limits for this quantity, computed
from Eq.~(\ref{eq:production}) with the same value of $\gammaz$
as used above.  Values for magnitude and phase of the \swave\
amplitude, and their statistical uncertainties, are computed
from Eq.~(\ref{eq:swave}) with parameters and error matrix 
from this fit.
The behaviour of $|\pvec_0(\SA)|$ derived from the isobar
model fit matches that observed in the \EIPWA\ points well.

The inset in Fig.~\ref{fig:smag_pvec} shows the corresponding 
quantities $|T_0(\SA)|/\sin(\gamma_{\sst B}+\gamma_{\sst R})$ 
for the points measured for $\Km\pip$ scattering in the LASS 
experiment.  From Eqs.~(\ref{eq:unitarity}) and 
(\ref{eq:lassdelta}) it is seen that this is expected, in the
range up to $\Km\eta^{\prime}$ threshold, to be unity.
It is seen that this condition is met by the LASS data.

It is concluded that the factor $|\pvec_0(\SA)|$ in
Eq.~(\ref{eq:production}) that describes production (and possible
re-scattering) for $\Km\pip$ systems in the $\Dp$ decays examined 
here, shows little dependence on \SA\ up to about 1.25~\GeVcc.  At 
this point, a significant dependence on \SA\ is seen.  This behaviour 
is qualitatively different from elastic scattering.


%% file: systematic.tex
The major source of systematic uncertainty in the \EIPWA\ results
arises from the difficulty, with a sample of this size, of reliably
characterizing the structures, other than the \KAp\ resonance, in the 
reference waves.  To estimate this effect, a large number of samples 
of MC events, each of which is of a size similar to the data 
($\sim 15K$ events) reported here, are examined.  These are 
generated with the parameters determined by the isobar model fit
described in Sec.~\ref{sec:kappatest}, with the
backgrounds best matched to the E791 data.  Each sample is 
subjected to a \EIPWA\ fit, and the differences between
generated and fitted values for \swave\ magnitude and phase at
each of the 40 invariant masses are examined.  For most samples, fits
obtained match the isobar model well.  Variations in the
significance of the \KCp\ and \KAd\ sometimes lead to variations 
in the reference waves that propagate to distortions in the \swave\
solutions found.
These tests provide estimates of systematic uncertainties for
the \swave\ magnitudes that range from $\sim~50$\% of the statistical
uncertainty, for invariant masses below 800~\MeVcc, to an 
insignificant level for higher masses.  For the \swave\ phases, the
systematic uncertainties are found to average $\sim 72$\%
of the statistical uncertainty.

The second largest uncertainty arises from the smearing of events
near the high mass boundary of the Dalitz plot which results from
the resolution in three-body mass $M$.  This directly affects part
of the \KCp\ band.  Events in the region of $M$ closest to the
$\Dp$ mass are fitted separately, and the results compared with that
from the larger sample.  Average systematic uncertainties arising
from the effects of smearing are determined to be 7\% of the
statistical uncertainties for magnitudes
and 14\% of the statistical uncertainties for phases.  Other
effects are studied.  These include the uncertainty in precise 
knowledge of the background level, variations in the values
assumed for the radii $\rres$ and $\rdpl$, or for the mass and width 
for the \KCp\ resonance.  All these other effects are found to be
small.

A further source of systematic uncertainty arises from variations
in the presumed resonant composition of the $\Km\pip$ \pdash\ and 
\dwave s.
The \KAp\ resonance obviously contributes, and it is clear that a
contribution from a higher resonance, must also exist.  What is less 
clear is the identity of this resonance - \KAd, \KCp\ or for \KBp.  
Fits are made with various combinations of these resonances.
It is found that systematic shifts are negligibly small in most cases.
Fits where only \KAp\ and \KAd\ are included do lead to shifts
comparable to the statistical uncertainties in the lowest
five magnitudes.  At higher invariant masses, the effects become smaller.
The phases are almost unchanged, however.
 
These uncertainties are combined in quadrature and listed for each
invariant mass in Table~\ref{tab:pwa}, and for the reference wave 
parameters in the \EIPWA\ fit in Table~\ref{tab:bigtab}.

%% file: summary.tex
A Model-Independent Partial Wave Analysis (\EIPWA)
of the \swave\ $\Km\pip$ system is made using the 
three body decay $\Dp\to\Km\pip\pip$.  This is the first time 
such a technique has been used in studying heavy quark meson 
decays, and new information on the $\Km\pip$ system is obtained,
including the invariant mass range below 825~\MeVcc.
The isospin $I$ of the \swave\ measured is unknown, and the
\pdash\ and \dwave s are assumed to be $I=\half$.  It is possible
to modify these assumptions, provided independent information
on the $I={3\over 2}$ components is available.
The method does not assume any form for the energy 
dependence of the \swave.  However, it does so for the \pdash\ and
\ddash\ reference waves.  The \pwave\ is described as the sum of a 
Breit-Wigner
propagator term for the \KAp\ resonance, and a similar term,
with a complex coefficient, for the \KCp.  The \KBp\ is found to
have an insignificant contribution to the decays, and is omitted
from this wave.  The \dwave\ is described by a single
Breit-Wigner term for the \KAd\ resonance, with a further complex
coefficient.  The results obtained
in Fig.~\ref{fig:solution0} and Table~\ref{tab:pwa}
depend on the accuracy of this description of the reference
waves.

Results of the \EIPWA\ are compared with a description of
the \swave\ amplitude that includes Breit-Wigner \KAs, \KBs\ 
isobars and a constant, non-resonant ($NR$) term similar to 
the description used in
Ref.~\cite{Aitala:2002kr}.  
At the statistical level of this experiment, differences between the 
\EIPWA\ and the isobar-model result are not found to be significant,
and both provide good descriptions of the data.
A closer examination of the phase behavior
in the low mass region below $825$~\MeVcc, the limit of measurements
of $\Km\pip$ elastic scattering from the LASS experiment
\cite{Aston:1987ir},
is of considerable importance to the further understanding of scalar
spectroscopy.
The data here provide new information in this region, but the
error bars are large compared to those typical for the LASS data.
We note that, since these data became available, a fit that includes
requirements of chiral 
perturbation theory has been made together with the LASS 
$I=1/2$ measurements and data from
$J/\psi\to\KAp\Km\pip$.  This fit finds a $\kappa$
pole at $(740^{+30}_{-55})-i(342\pm 60)$~\MeVcc
\cite{Bugg:2005xx}.
A full understanding of scalar K* spectroscopy may, nevertheless,
need to wait until larger data samples become available.
A better consensus on the proper theoretical description of such
states and the need for, and the form of,
any accompanying background amplitudes may also be required.

The phases observed in the \sdash\ and \pwave s do not appear
to match those seen in the $I=\half$ elastic scattering in reference
\cite{Aston:1987ir}.  The \dwave\ phase does agree well.  
Constraining the energy dependence
of the \swave\ phase to follow that observed in $(I=\half)$ $K\pi$ 
elastic scattering, in the range where \SA\ lies below $K\eta'$ 
threshold, does lead to a good fit to the data.
However, an overall shift in phase of $(-74.4\pm 1.8\pm 1.0)^{\circ}$ 
relative to the \pwave\ is still required.  This constraint also
results in a shift of approximately $-50^{\circ}$ in the \dwave\ phase.
It also makes agreement in \pwave\ phase worse.
These results do not conform, exactly, to the expectations of the
Watson theorem
which would require phases in each wave to match, apart from an overall
shift, those for $\Km\pip$
scattering for invariant masses below $K\eta'$ threshold.  The theorem
is expected to apply in kinematic regions where secondary scattering
of the $K\pi$ system from the bachelor pion can be neglected.  It is
possible that, in this case, such scattering cannot be neglected, or
that the $\Km\pip$ systems in $\Dp$ decay are not predominantly
$(I=\half)$
\cite{Edera:2005dk}.

It is also found that, with a choice of phase at $\Km\pip$ threshold 
relative to the \pwave\ $\gammaz = (-123.3 \pm 3.9)^{\circ}$,
quite consistent with that measured in the \EIPWA,
$\Km\pip$ systems produced from $\Dp$ decays are
described well by a unitary amplitude (with constant production) up to
a mass of about 1.25~\GeVcc.  In this region, therefore,
structure observed in the \swave\ magnitude
is mainly associated with the variation in phase with respect to
\SA. the invariant mass squared in the $\Km\pip$ system.  Above 
1.25~\GeVcc, the production rate grows, depending significantly on 
\SA.  The reason for this behavior is unknown.  The growth observed
at 1.25~\GeVcc\ could result from a significant
$I={3\over 2}$ contribution or from re-scattering of the produced
$\Km\pip$ system and the bachelor $\pip$.

The \EIPWA\ analysis of the three-body decay of a heavy quark system
described here has three main limitations.  The first results from the
way the reference \pwave\ is described in Eq.~(\ref{eq:pwave}).
Using Breit-Wigner resonance forms for more than one resonance in the
wave can lead to problems in the regions where the resonance tails 
dominate.
The second limitation comes from the ability to resolve the
structure in the \KCp\ region properly at the statistical level of
the E791 data.  This problem may be specific to the channel discussed
here, and to the particular data sample used.  The third limitation
is the lack of knowledge on any $I={3\over 2}$ components in the
system.

The first two limitations should be mitigated when much larger data 
samples are available.  A better formulation for the \pwave\
could be to use a $K$-matrix, requiring more parameters.  Alternatively,
the \pwave, too, could be parametrized like the \swave, in a
model-independent way.  The third limitation can be improved when
larger samples of $\Kp\pip$ or $\Km\pim$ systems can be studied to
better understand these waves.

Systematic studies of various heavy quark meson decays in future 
experiments (BaBar, Belle, CLEO-c, and hadron colliders), with 
much larger samples, may be able to use a similar \EIPWA\ technique,
with some of these improvements, to shed further light on important 
questions in light quark spectroscopy, the realm of applicability of 
the Watson theorem, {\sl etc}.  For studies that require an empirically
good description of the complex amplitude in three body decays, for 
example in the extraction of the $\gamma$ CP violation parameter 
recently reported by BaBar and Belle
\cite{Aubert:2005iz, Poluektov:2004mf, Abe:2005ct},
this technique may also be particularly useful.

In the mean time, theoretical models of the S-wave amplitude can
be compared to the data of Table III.

%% file: ambiguities.tex
\label{sec:pwave}
\newcommand{\swa}{\ensuremath{{\cal S}_{\sst A}}}
\newcommand{\pwa}{\ensuremath{{\cal P}_{\sst A}}}
\newcommand{\dwa}{\ensuremath{{\cal D}_{\sst A}}}
\newcommand{\swb}{\ensuremath{{\cal S}_{\sst B}}}
\newcommand{\pwb}{\ensuremath{{\cal P}_{\sst B}}}
\newcommand{\dwb}{\ensuremath{{\cal D}_{\sst B}}}
\newcommand{\twa}{\ensuremath{{\cal T}_{\sst A}}}
\subsection{Quality of \swave\ Measurements}
\label{sec:sw_measure}
\begin{figure*}[hbt]
 \centerline{%
 \epsfig{file=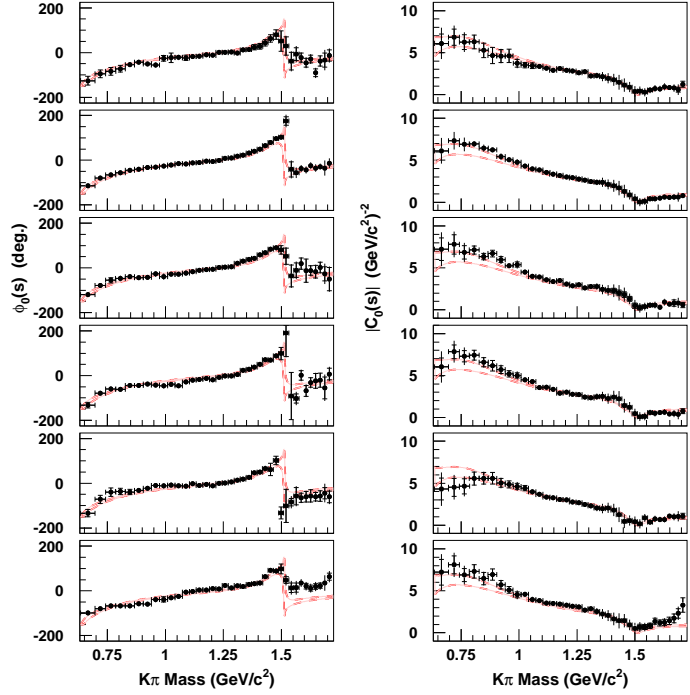,height=0.45\textheight,angle=0}}
 \caption{Comparison between fitted \swave\ amplitudes (solid
  circles with error bars representing statistical uncertainties)
  and generated amplitudes, represented as shaded regions between
  dashed curves computed as described in Sec.~\ref{sec:kappatest}.
  Generated curves follow the isobar fit model described in 
  Sec.~\ref{sec:kappatest}.  The figure shows results from the
  first six (of 100) samples used in the test.
 \label{fig:mc15k}}
\end{figure*}
The \EIPWA\ analysis of the \swave\ component in the observed
$\Dp\to\Km\piA\piB$ decays relies upon a good description of
the reference \pdash\ and \dwave s.

The \pwave\ defined in Eq.~(\ref{eq:pwave}) is a combination
of two Breit-Wigner's (the \KBp\ is neglected), 
with a complex coupling coefficient (two parameters).
The peak regions, $0.50<\SA<0.9$~(\GeVcc)$^2$ and $\SA>2.2$
~(\GeVcc)$^2$,
are well described by this parametrization, since data in 
these regions are likely to be dominated by these resonances.
In the tail regions, the Breit-Wigner may be a less appropriate 
description of the data, since other \pwave\ contributions, 
from non-resonant (or $I={3\over 2}$) sources, for example,
could become more significant.

Two regions in the \pwave\ where the tails of the BW's dominate
are $\SA<0.50$~(\GeVcc)$^2$ and $0.9<\SA<2.2$~(\GeVcc)$^2$.  In 
each of these ranges the \pwave\ is constructed from a linear 
combination of two small, complex numbers, one from each of 
the two BW tails.  Both the phase and magnitude of the 
resultant are particularly sensitive to variations in the 
complex coupling parameters, and may not represent the \pwave\
well.

The \dwave\ is defined in Eq.~(\ref{eq:dwave}), for this analysis,
as a single, $L=2$ BW function.  Non-resonant contributions
are not expected to be significant, and interference from
tails of a second resonance are absent.  Eq.~(\ref{eq:dwave}),
therefore, provides a relatively good description of the \dwave.

Eqs.~(\ref{eq:pwa_bose}) and 
(\ref{eq:pwa_expand}) show that the amplitude for the decays 
examined in this paper is a sum of six terms.  Let these be 
labelled \swa, \pwa\
and \dwa\ (the \sdash, \pdash\ and \dwave, respectively, in 
the \SAA\ channel) and \swb, \pwb\ and \dwb\ (these waves in
the \SAB\ channel).  
The \EIPWA\ process extracts magnitude and phase information
about the \swave\ \swa\ from its observed
interference with the complex sum of the other five amplitudes.
\bea
  \label{eq:analyzer}
    \twa = \pwa+\dwa+\swb+\pwb+\dwb.
\eea
%
The results expected from measurement of \swa\ can, therefore,
be characterized by the dominant terms in \twa\ with which
it interferes, and these depend on location on the Dalitz plot
in Fig.~\ref{fig:dalitz_plot}.  

As an illustration, consider
measurement of \swa\ in the range $1.1<\SA<2.9$~(\GeVcc)$^2$.
Here, \twa\ is dominated by the \KAp\ resonance band in \pwb\
(the cross-channel).  Good measurements are, therefore,
expected in this range.

Next, consider the \KAp\ peak region $1.5<\SA<0.9$~(\GeVcc)$^2$.  
This region has \twa\ dominated by the \KAp\ peak in \pwa\ (the
direct-channel).  So good measurements are expected here too.
A similar conclusion can be drawn for the \KCp\ peak region
$\SA>2.9$~(\GeVcc)$^2$.

Relatively poor measurements are expected for the other
regions since, in these, \twa\ is not dominated by any one 
source, and is defined by a linear combination of several
Breit-Wigner tails.  So \twa\ in these regions has phase and 
magnitude that are sensitive to the complex couplings of \KCp\ 
and \KAd\ resonances.

These observations are supported by the results of the \EIPWA\
fit shown in Fig.~\ref{fig:solution0}(a) and (b).  It is
seen that uncertainties are small for $1.1<\SA<2.9$~(\GeVcc)$^2$
and large for $\SA<0.9$~(\GeVcc)$^2$, improving towards the
high end.  In the intermediate
region, $0.9<\SA<1.1$~(\GeVcc)$^2$, the \swave\ magnitudes and
phases determined in the fit exhibit significant
deviations from the general trends of the neighboring points.

\subsection{MC Studies with Isobar Fit}
\label{sec:mcstudies}
The \KCp\ and \KAd\ resonances represent small contributions to 
the Dalitz plot, and statistical uncertainties in their complex
couplings are large enough to affect the \pwave\ phase, 
especially in the region between \KAp\ and \KCp, as discussed in 
Sec.~\ref{sec:sw_measure}.
This can lead to systematic uncertainties in the \swave\ 
amplitudes measured.  MC studies are required to estimate 
such effects.

MC samples of the approximate size of the data presented in 
this paper are generated as described by the PDF given in
Eq.~(\ref{eq:pdfsig}).  Parameters from Table~\ref{tab:bigtab}
for the isobar fit described in Sec.~\ref{sec:kappatest} are
used for this purpose.  Background events whose distributions
are given in Eq.~(\ref{eq:pdfback}) are also generated to
match those thought to be present in the data.  Events are
selected according to the efficiency $\epsilon(\SAA,\SAB)$
across the Dalitz plot.

Each sample is subjected to the \EIPWA\ fit described in 
Sec.~\ref{sec:swave}.  
In Fig.~\ref{fig:mc15k}, \swave\ amplitudes determined in the 
\EIPWA\ for the first six of the 100 samples studied are 
compared with
those used to generate the events.  The amplitudes generated
come from the isobar fit, and are shown, as usual, as shaded 
regions between dashed curves.  Phases are shown on the right
and magnitudes on the left.  Plots similar to 
Fig.~\ref{fig:solution0} appear often.

The \swave\ amplitudes ($c_k e^{i\gamma_k}$) obtained are compared 
with those generated and, for each $k$ the normalized residuals 
are used to determine systematic uncertainties discussed in 
Sec.~\ref{sec:systematic}.

\subsection{\bf Other Solutions}
\begin{figure*}[hbt]
 \centerline{%
 \epsfig{file=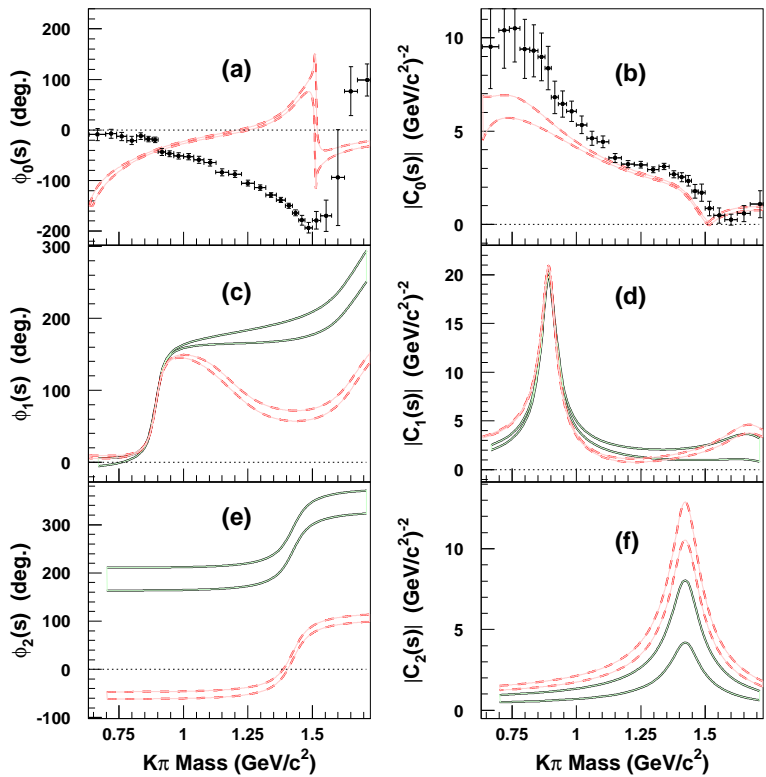,height=0.5\textheight,angle=0}}
 \caption{A second solution for the \swave\ amplitude from \EIPWA\
  fits to $\Dp\to\Km\pip\pip$ decays with \pdash\ and \dwave\
  parametrized by the $\kappa$ model described in the text.  Plots
  show the 
  (a) phase and (b) magnitude for solution B for the \swave\ 
  obtained by using
  different starting values for the amplitudes.  The dashed curves
  delineate the regions that lie within one standard deviation of 
  the isobar model fit described in Sec.~\ref{sec:kappatest}.  The 
  \pwave\ is shown in (c) and (d) and the \dwave\ in (e) and (f).
 \label{fig:solnb}}
\end{figure*}
The fitting procedure allows a great deal of freedom to the \swave\
amplitude.  Consequently, ambiguities in solutions are
anticipated.  To study possible ambiguities in the \EIPWA\ solution,
fits with random starting values for the $c_i, \gamma_i$ 
parameters, and also with different $\Km\pip$ mass slices are made.
One other local maximum in the likelihood is found, and this is
labelled solution B.  The solution described in 
Sec.~\ref{sec:swave}, and shown in Figs.~\ref{fig:solution0}(a) 
through (f), is labelled, for contrast, solution A.  Solution A is the 
only one with an acceptable $\chidof$ and has the greatest 
likelihood value.  So it is emphasized that solutions A is, in fact,
unique.


Solution B is shown in Fig.~\ref{fig:solnb}.  It provides a qualitatively
reasonable description of the distribution of the data on the Dalitz plot.
However, this solution clearly exhibits retrograde motion around the
unitarity circle as $\Km\pip$ invariant mass increases.
This violates the Wigner causality principle
\cite{Wigner:1955ca},
thus eliminating it from further consideration.

The possible existence of other maxima in the likelihood, when all
\swave\ magnitudes and phases are free parameters, cannot be
completely ruled out.  However, the solution in Sec.~\ref{sec:swave} 
is unique in that it is the only one giving an acceptable fit 
probability.

%% file: preprint.bbl
\begin{thebibliography}{36}
\expandafter\ifx\csname natexlab\endcsname\relax\def\natexlab#1{#1}\fi
\expandafter\ifx\csname bibnamefont\endcsname\relax
  \def\bibnamefont#1{#1}\fi
\expandafter\ifx\csname bibfnamefont\endcsname\relax
  \def\bibfnamefont#1{#1}\fi
\expandafter\ifx\csname citenamefont\endcsname\relax
  \def\citenamefont#1{#1}\fi
\expandafter\ifx\csname url\endcsname\relax
  \def\url#1{\texttt{#1}}\fi
\expandafter\ifx\csname urlprefix\endcsname\relax\def\urlprefix{URL }\fi
\providecommand{\bibinfo}[2]{#2}
\providecommand{\eprint}[2][]{\url{#2}}

\bibitem[{con()}]{conjugate}
\bibinfo{note}{{Charge conjugate states are always implied unless explicitly
  stated otherwise.}}

\bibitem[{\citenamefont{Frabetti et~al.}(1994)}]{Frabetti:1994di}
\bibinfo{author}{\bibfnamefont{P.~L.} \bibnamefont{Frabetti}}
  \bibnamefont{et~al.} (\bibinfo{collaboration}{E687}), \bibinfo{journal}{Phys.
  Lett.} \textbf{\bibinfo{volume}{B331}}, \bibinfo{pages}{217}
  (\bibinfo{year}{1994}).

\bibitem[{\citenamefont{Anjos et~al.}(1993)}]{Anjos:1992kb}
\bibinfo{author}{\bibfnamefont{J.~C.} \bibnamefont{Anjos}} \bibnamefont{et~al.}
  (\bibinfo{collaboration}{E691}), \bibinfo{journal}{Phys. Rev.}
  \textbf{\bibinfo{volume}{D48}}, \bibinfo{pages}{56} (\bibinfo{year}{1993}).

\bibitem[{\citenamefont{Frabetti et~al.}(1997)}]{Frabetti:1997sx}
\bibinfo{author}{\bibfnamefont{P.~L.} \bibnamefont{Frabetti}}
  \bibnamefont{et~al.} (\bibinfo{collaboration}{E687}), \bibinfo{journal}{Phys.
  Lett.} \textbf{\bibinfo{volume}{B407}}, \bibinfo{pages}{79}
  (\bibinfo{year}{1997}).

\bibitem[{\citenamefont{Aitala et~al.}(2001)}]{Aitala:2000xu}
\bibinfo{author}{\bibfnamefont{E.~M.} \bibnamefont{Aitala}}
  \bibnamefont{et~al.} (\bibinfo{collaboration}{E791}), \bibinfo{journal}{Phys.
  Rev. Lett.} \textbf{\bibinfo{volume}{86}}, \bibinfo{pages}{770}
  (\bibinfo{year}{2001}), \eprint{hep-ex/0007028}.

\bibitem[{\citenamefont{Aitala et~al.}(2002)}]{Aitala:2002kr}
\bibinfo{author}{\bibfnamefont{E.~M.} \bibnamefont{Aitala}}
  \bibnamefont{et~al.} (\bibinfo{collaboration}{E791}), \bibinfo{journal}{Phys.
  Rev. Lett.} \textbf{\bibinfo{volume}{89}}, \bibinfo{pages}{121801}
  (\bibinfo{year}{2002}), \eprint{hep-ex/0204018}.

\bibitem[{\citenamefont{Link et~al.}(2004)}]{Link:2003gb}
\bibinfo{author}{\bibfnamefont{J.~M.} \bibnamefont{Link}} \bibnamefont{et~al.}
  (\bibinfo{collaboration}{FOCUS}), \bibinfo{journal}{Phys. Lett.}
  \textbf{\bibinfo{volume}{B585}}, \bibinfo{pages}{200} (\bibinfo{year}{2004}),
  \eprint{hep-ex/0312040}.

\bibitem[{\citenamefont{Aubert et~al.}(2005{\natexlab{a}})}]{Aubert:2005iz}
\bibinfo{author}{\bibfnamefont{B.}~\bibnamefont{Aubert}} \bibnamefont{et~al.}
  (\bibinfo{collaboration}{BABAR}) (\bibinfo{year}{2005}{\natexlab{a}}),
  \bibinfo{note}{{Submitted to \PRL.}}, \eprint{hep-ex/0504039}.

\bibitem[{\citenamefont{Poluektov et~al.}(2004)}]{Poluektov:2004mf}
\bibinfo{author}{\bibfnamefont{A.}~\bibnamefont{Poluektov}}
  \bibnamefont{et~al.} (\bibinfo{collaboration}{Belle}),
  \bibinfo{journal}{Phys. Rev.} \textbf{\bibinfo{volume}{D70}},
  \bibinfo{pages}{072003} (\bibinfo{year}{2004}), \eprint{hep-ex/0406067}.

\bibitem[{\citenamefont{Abe et~al.}(2005)}]{Abe:2005ct}
\bibinfo{author}{\bibfnamefont{K.}~\bibnamefont{Abe}} \bibnamefont{et~al.}
  (\bibinfo{collaboration}{Belle}) (\bibinfo{year}{2005}),
  \bibinfo{note}{{Contributed to 40th Rencontres de Moriond on Electroweak
  Interactions and Unified Theories, La Thuile, Aosta Valley, Italy, 5-12 Mar
  2005.}}, \eprint{hep-ex/0504013}.

\bibitem[{\citenamefont{Ablikim et~al.}(2004)}]{Ablikim:2004qn}
\bibinfo{author}{\bibfnamefont{M.}~\bibnamefont{Ablikim}} \bibnamefont{et~al.}
  (\bibinfo{collaboration}{BES}), \bibinfo{journal}{Phys. Lett.}
  \textbf{\bibinfo{volume}{B598}}, \bibinfo{pages}{149} (\bibinfo{year}{2004}),
  \eprint{hep-ex/0406038}.

\bibitem[{\citenamefont{Komada}(2004)}]{Komada:2003gu}
\bibinfo{author}{\bibfnamefont{T.}~\bibnamefont{Komada}}, \bibinfo{journal}{AIP
  Conf. Proc.} \textbf{\bibinfo{volume}{717}}, \bibinfo{pages}{337}
  (\bibinfo{year}{2004}).

\bibitem[{\citenamefont{Gardner and Meissner}(2002)}]{Gardner:2001gc}
\bibinfo{author}{\bibfnamefont{S.}~\bibnamefont{Gardner}} \bibnamefont{and}
  \bibinfo{author}{\bibfnamefont{U.-G.} \bibnamefont{Meissner}},
  \bibinfo{journal}{Phys. Rev.} \textbf{\bibinfo{volume}{D65}},
  \bibinfo{pages}{094004} (\bibinfo{year}{2002}), \eprint{hep-ph/0112281}.

\bibitem[{\citenamefont{Oller}(2005)}]{Oller:2004xm}
\bibinfo{author}{\bibfnamefont{J.~A.} \bibnamefont{Oller}},
  \bibinfo{journal}{Phys. Rev.} \textbf{\bibinfo{volume}{D71}},
  \bibinfo{pages}{054030} (\bibinfo{year}{2005}), \eprint{hep-ph/0411105}.

\bibitem[{\citenamefont{Oller and Oset}(1997)}]{Oller:1997ti}
\bibinfo{author}{\bibfnamefont{J.~A.} \bibnamefont{Oller}} \bibnamefont{and}
  \bibinfo{author}{\bibfnamefont{E.}~\bibnamefont{Oset}},
  \bibinfo{journal}{Nucl. Phys.} \textbf{\bibinfo{volume}{A620}},
  \bibinfo{pages}{438} (\bibinfo{year}{1997}), \eprint{hep-ph/9702314}.

\bibitem[{\citenamefont{Jamin et~al.}(2000)\citenamefont{Jamin, Oller, and
  Pich}}]{Jamin:2000wn}
\bibinfo{author}{\bibfnamefont{M.}~\bibnamefont{Jamin}},
  \bibinfo{author}{\bibfnamefont{J.~A.} \bibnamefont{Oller}}, \bibnamefont{and}
  \bibinfo{author}{\bibfnamefont{A.}~\bibnamefont{Pich}},
  \bibinfo{journal}{Nucl. Phys.} \textbf{\bibinfo{volume}{B587}},
  \bibinfo{pages}{331} (\bibinfo{year}{2000}), \eprint{hep-ph/0006045}.

\bibitem[{\citenamefont{Bediaga and de~Miranda}(2004)}]{Bediaga:2004bc}
\bibinfo{author}{\bibfnamefont{I.}~\bibnamefont{Bediaga}} \bibnamefont{and}
  \bibinfo{author}{\bibfnamefont{J.~M.} \bibnamefont{de~Miranda}}
  (\bibinfo{year}{2004}), \bibinfo{note}{{ Submitted to \PLB.}},
  \eprint{hep-ex/0405019}.

\bibitem[{\citenamefont{Bediaga and de~Miranda}(2002)}]{Bediaga:2002au}
\bibinfo{author}{\bibfnamefont{I.}~\bibnamefont{Bediaga}} \bibnamefont{and}
  \bibinfo{author}{\bibfnamefont{J.~M.} \bibnamefont{de~Miranda}},
  \bibinfo{journal}{Phys. Lett.} \textbf{\bibinfo{volume}{B550}},
  \bibinfo{pages}{135} (\bibinfo{year}{2002}), \eprint{hep-ph/0211078}.

\bibitem[{\citenamefont{Aston et~al.}(1988)}]{Aston:1987ir}
\bibinfo{author}{\bibfnamefont{D.}~\bibnamefont{Aston}} \bibnamefont{et~al.}
  (\bibinfo{collaboration}{LASS}), \bibinfo{journal}{Nucl. Phys.}
  \textbf{\bibinfo{volume}{B296}}, \bibinfo{pages}{493} (\bibinfo{year}{1988}).

\bibitem[{\citenamefont{Bingham et~al.}(1972)}]{Bingham:1972vy}
\bibinfo{author}{\bibfnamefont{H.~H.} \bibnamefont{Bingham}}
  \bibnamefont{et~al.}, \bibinfo{journal}{Nucl. Phys.}
  \textbf{\bibinfo{volume}{B41}}, \bibinfo{pages}{1} (\bibinfo{year}{1972}).

\bibitem[{\citenamefont{Estabrooks et~al.}(1978)}]{Estabrooks:1977xe}
\bibinfo{author}{\bibfnamefont{P.}~\bibnamefont{Estabrooks}}
  \bibnamefont{et~al.}, \bibinfo{journal}{Nucl. Phys.}
  \textbf{\bibinfo{volume}{B133}}, \bibinfo{pages}{490} (\bibinfo{year}{1978}).

\bibitem[{\citenamefont{Aubert et~al.}(2005{\natexlab{b}})}]{Aubert:2004cp}
\bibinfo{author}{\bibfnamefont{B.}~\bibnamefont{Aubert}} \bibnamefont{et~al.}
  (\bibinfo{collaboration}{BABAR}), \bibinfo{journal}{Phys. Rev.}
  \textbf{\bibinfo{volume}{D71}}, \bibinfo{pages}{032005}
  (\bibinfo{year}{2005}{\natexlab{b}}), \eprint{hep-ex/0411016}.

\bibitem[{\citenamefont{Link et~al.}(2002)}]{Link:2002ev}
\bibinfo{author}{\bibfnamefont{J.~M.} \bibnamefont{Link}} \bibnamefont{et~al.}
  (\bibinfo{collaboration}{FOCUS}), \bibinfo{journal}{Phys. Lett.}
  \textbf{\bibinfo{volume}{B535}}, \bibinfo{pages}{43} (\bibinfo{year}{2002}),
  \eprint{hep-ex/0203031}.

\bibitem[{\citenamefont{Watson}(1952)}]{watson:1952ji}
\bibinfo{author}{\bibfnamefont{K.~M.} \bibnamefont{Watson}},
  \bibinfo{journal}{Phys. Rev.} \textbf{\bibinfo{volume}{88}},
  \bibinfo{pages}{1163} (\bibinfo{year}{1952}).

\bibitem[{\citenamefont{Aitala et~al.}(1999)}]{Aitala:1998kh}
\bibinfo{author}{\bibfnamefont{E.~M.} \bibnamefont{Aitala}}
  \bibnamefont{et~al.} (\bibinfo{collaboration}{E791}), \bibinfo{journal}{Eur.
  Phys. J. direct} \textbf{\bibinfo{volume}{C1}}, \bibinfo{pages}{4}
  (\bibinfo{year}{1999}), \eprint[http://arXiv.org/abs]{hep-ex/9809029}.

\bibitem[{hel()}]{helicity}
\bibinfo{note}{{This definition of helicity angle differs from that used in
  $\Km\pip$ scattering where the angle is defined as that btween $\Km$ and the
  momentum of the $\Km\pip$ system.}}

\bibitem[{\citenamefont{Hoogland et~al.}(1977)}]{Hoogland:1977kt}
\bibinfo{author}{\bibfnamefont{W.}~\bibnamefont{Hoogland}}
  \bibnamefont{et~al.}, \bibinfo{journal}{Nucl. Phys.}
  \textbf{\bibinfo{volume}{B126}}, \bibinfo{pages}{109} (\bibinfo{year}{1977}).

\bibitem[{\citenamefont{Blatt and Weisskopf}(1952)}]{BlattWeiss}
\bibinfo{author}{\bibfnamefont{J.~M.} \bibnamefont{Blatt}} \bibnamefont{and}
  \bibinfo{author}{\bibfnamefont{V.~F.} \bibnamefont{Weisskopf}},
  \bibinfo{journal}{Wiley, New York} p. \bibinfo{pages}{361}
  (\bibinfo{year}{1952}).

\bibitem[{\citenamefont{Tornqvist}(1995)}]{Tornqvist:1995kr}
\bibinfo{author}{\bibfnamefont{N.~A.} \bibnamefont{Tornqvist}},
  \bibinfo{journal}{Z. Phys.} \textbf{\bibinfo{volume}{C68}},
  \bibinfo{pages}{647} (\bibinfo{year}{1995}), \eprint{hep-ph/9504372}.

\bibitem[{Adl()}]{Adler}
\bibinfo{note}{{It has often been suggested that the denominator should include
  a factor $\SA-s_A$ where $s_A=m_K^2-0.5 m_{\pi}^2$ is the location of the
  Adler zero required in $\Km\pip$ scattering.}}

\bibitem[{Spl()}]{Spline}
\bibinfo{note}{{To avoid difficulties arising from the periodic nature of the
  phase $\gamma_k$, the interpolation is made in the values of the real and
  imaginary parts of the amplitude $c_ke^{i\gamma_k}$. For each of these,
  quadratic functions are defined to the left and right of each \SAk\ such that
  their first derivatives are equal at each point.}}

\bibitem[{\citenamefont{{S. Eidelman et al.}}(2004)}]{PDG}
\bibinfo{author}{\bibnamefont{{S. Eidelman et al.}}}
  (\bibinfo{collaboration}{{PDG}}), \bibinfo{journal}{{Physics Letters B}}
  \textbf{\bibinfo{volume}{592}}, \bibinfo{pages}{1} (\bibinfo{year}{2004}),
  \urlprefix\url{http://pdg.lbl.gov}.

\bibitem[{loc()}]{localmin}
\bibinfo{note}{{Other local minima are also found. These are discussed in
  Appendix~\ref{sec:ambiguities}. The fit solutions described here give the
  greatest likelihood value and, based on $\chidof$, the best description of
  the data.}}

\bibitem[{\citenamefont{Bugg}(2005)}]{Bugg:2005xx}
\bibinfo{author}{\bibfnamefont{D.~V.} \bibnamefont{Bugg}}
  (\bibinfo{year}{2005}), \eprint{hep-ex/0510019}.

\bibitem[{\citenamefont{Edera and Pennington}(2005)}]{Edera:2005dk}
\bibinfo{author}{\bibfnamefont{L.}~\bibnamefont{Edera}} \bibnamefont{and}
  \bibinfo{author}{\bibfnamefont{M.~R.} \bibnamefont{Pennington}}
  (\bibinfo{year}{2005}), \bibinfo{note}{{have recently elaborated on this
  point.}}, \eprint{hep-ph/0506117}.

\bibitem[{\citenamefont{Wigner}(1955)}]{Wigner:1955ca}
\bibinfo{author}{\bibfnamefont{E.~P.} \bibnamefont{Wigner}},
  \bibinfo{journal}{Phys. Rev.} \textbf{\bibinfo{volume}{98}},
  \bibinfo{pages}{145} (\bibinfo{year}{1955}).

\end{thebibliography}
